\UseRawInputEncoding

\documentclass[twocolumn]{aastex63}

\usepackage{amsmath}
\usepackage[super]{nth}
\usepackage{enumitem}

\received{March 9, 2021}
\revised{March 26, 2021}
\accepted{March 29, 2021}

\shorttitle{BISTRO-2 Survey: Magnetic Field in the RMC}
\shortauthors{K\"onyves et al.}

\begin{document}

\title{The JCMT BISTRO-2 Survey: The Magnetic Field in the Center of the Rosette Molecular Cloud}

\correspondingauthor{Vera K\"onyves}
\email{vkonyves@uclan.ac.uk}

\author[0000-0002-3746-1498]{Vera K\"{o}nyves}
\affiliation{Jeremiah Horrocks Institute, University of Central Lancashire, Preston PR1 2HE, UK}

\author[0000-0003-1140-2761]{Derek Ward-Thompson}
\affiliation{Jeremiah Horrocks Institute, University of Central Lancashire, Preston PR1 2HE, UK}

\author[0000-0002-8557-3582]{Kate Pattle}
\affiliation{Centre for Astronomy, School of Physics, National University of Ireland Galway, University Road, Galway H91 TK33, Ireland}

\author[0000-0002-9289-2450]{James Di Francesco}
\affiliation{NRC Herzberg Astronomy and Astrophysics, 5071 West Saanich Road, Victoria, BC V9E 2E7, Canada}
\affiliation{Department of Physics and Astronomy, University of Victoria, Victoria, BC V8W 2Y2, Canada}

\author[0000-0002-1959-7201]{Doris Arzoumanian}
\affiliation{Aix Marseille Univ., CNRS, CNES, LAM, Marseille, France}

\author[0000-0003-0849-0692]{Zhiwei Chen}
\affiliation{Purple Mountain Observatory, Chinese Academy of Sciences, 10 Yuanhua, 210033 Nanjing, People's Republic of China}

\author[0000-0002-2808-0888]{Pham Ngoc Diep}
\affiliation{Vietnam National Space Center, Vietnam Academy of Science and Technology, 18 Hoang Quoc Viet, Hanoi, Vietnam}

\author[0000-0003-4761-6139]{Chakali Eswaraiah}
\affiliation{CAS Key Laboratory of FAST, National Astronomical Observatories, Chinese Academy of Sciences, Peopleʼs Republic of China}
\affiliation{National Astronomical Observatories, Chinese Academy of Sciences, A20 Datun Road, Chaoyang District, Beijing 100012, People’s Republic of China} 

\author[0000-0001-9930-9240]{Lapo Fanciullo}
\affiliation{Academia Sinica Institute of Astronomy and Astrophysics, No.1, Sec. 4., Roosevelt Road, Taipei 10617, Taiwan}

\author[0000-0003-0646-8782]{Ray S. Furuya}
\affiliation{Tokushima University, Minami Jousanajima-machi 1-1, Tokushima 770-8502, Japan}
\affiliation{Institute of Liberal Arts and Sciences Tokushima University, Minami Jousanajima-machi 1-1, Tokushima 770-8502, Japan}

\author[0000-0003-2017-0982]{Thiem Hoang}
\affiliation{Korea Astronomy and Space Science Institute, 776 Daedeokdae-ro, Yuseong-gu, Daejeon 34055, Republic of Korea}
\affiliation{University of Science and Technology, Korea, 217 Gajeong-ro, Yuseong-gu, Daejeon 34113, Republic of Korea}

\author[0000-0002-8975-7573]{Charles L. H. Hull}
\affiliation{National Astronomical Observatory of Japan, Alonso de C\'ordova 3788, Office 61B, Vitacura, Santiago, Chile}
\affiliation{Joint ALMA Observatory, Alonso de C\'ordova 3107, Vitacura, Santiago, Chile}		
\affiliation{NAOJ Fellow}		

\author[0000-0001-7866-2686]{Jihye Hwang}
\affiliation{Korea Astronomy and Space Science Institute, 776 Daedeokdae-ro, Yuseong-gu, Daejeon 34055, Republic of Korea}
\affiliation{University of Science and Technology, Korea, 217 Gajeong-ro, Yuseong-gu, Daejeon 34113, Republic of Korea}				

\author[0000-0002-6773-459X]{Doug Johnstone}
\affiliation{NRC Herzberg Astronomy and Astrophysics, 5071 West Saanich Road, Victoria, BC V9E 2E7, Canada}
\affiliation{Department of Physics and Astronomy, University of Victoria, Victoria, BC V8W 2Y2, Canada}

\author[0000-0001-7379-6263]{Ji-hyun Kang}
\affiliation{Korea Astronomy and Space Science Institute, 776 Daedeokdae-ro, Yuseong-gu, Daejeon 34055, Republic of Korea}

\author{Janik Karoly}
\affiliation{Jeremiah Horrocks Institute, University of Central Lancashire, Preston PR1 2HE, UK}

\author[0000-0002-3036-0184]{Florian Kirchschlager}
\affiliation{Department of Physics and Astronomy, University College London, WC1E 6BT London, UK}

\author[0000-0002-4552-7477]{Jason M. Kirk}
\affiliation{Jeremiah Horrocks Institute, University of Central Lancashire, Preston PR1 2HE, UK}

\author[0000-0003-2777-5861]{Patrick M. Koch}
\affiliation{Academia Sinica Institute of Astronomy and Astrophysics, No.1, Sec. 4., Roosevelt Road, Taipei 10617, Taiwan}

\author[0000-0003-2815-7774]{Jungmi Kwon}
\affiliation{Department of Astronomy, Graduate School of Science, The University of Tokyo, 7-3-1 Hongo, Bunkyo-ku, Tokyo 113-0033, Japan}

\author[0000-0002-3179-6334]{Chang Won Lee}
\affiliation{Korea Astronomy and Space Science Institute, 776 Daedeokdae-ro, Yuseong-gu, Daejeon 34055, Republic of Korea}
\affiliation{University of Science and Technology, Korea, 217 Gajeong-ro, Yuseong-gu, Daejeon 34113, Republic of Korea}

\author[0000-0002-8234-6747]{Takashi Onaka}
\affiliation{Department of Physics, Faculty of Science and Engineering, Meisei University, 2-1-1 Hodokubo, Hino, Tokyo 191-8506, Japan}
\affiliation{Department of Astronomy, Graduate School of Science, The University of Tokyo, 7-3-1 Hongo, Bunkyo-ku, Tokyo 113-0033, Japan}

\author[0000-0001-5079-8573]{Jean-Fran\c{c}ois Robitaille}
\affiliation{Univ. Grenoble Alpes, CNRS, IPAG, 38000 Grenoble, France}

\author[0000-0002-6386-2906]{Archana Soam}
\affiliation{SOFIA Science Center, Universities Space Research Association, NASA Ames Research Center, Moffett Field, California 94035, USA}

\author[0000-0001-8749-1436]{Mehrnoosh Tahani}
\affiliation{Dominion Radio Astrophysical Observatory, Herzberg Astronomy and Astrophysics Research Centre, National Research Council Canada, P. O. Box 248, Penticton, BC V2A 6J9 Canada}

\author[0000-0002-4154-4309]{Xindi Tang}
\affiliation{Xinjiang Astronomical Observatory, Chinese Academy of Sciences, 830011 Urumqi, Peopleʼs Republic of China}

\author[0000-0002-6510-0681]{Motohide Tamura}
\affiliation{National Astronomical Observatory of Japan, National Institutes of Natural Sciences, Osawa, Mitaka, Tokyo 181-8588, Japan} 
\affiliation{Department of Astronomy, Graduate School of Science, The University of Tokyo, 7-3-1 Hongo, Bunkyo-ku, Tokyo 113-0033, Japan}
\affiliation{Astrobiology Center, National Institutes of Natural Sciences, 2-21-1 Osawa, Mitaka, Tokyo 181-8588, Japan} 	

\author[0000-0001-6524-2447]{David Berry}
\affiliation{East Asian Observatory, 660 N. A'oh\={o}k\={u} Place, University Park, Hilo, HI 96720, USA}

\author[0000-0002-0794-3859]{Pierre Bastien}
\affiliation{Centre de recherche en astrophysique du Qu\'{e}bec \& d\'{e}partement de physique, Universit\'{e} de Montr\'{e}al, C.P. 6128 Succ. Centre-ville, Montr\'{e}al, QC, H3C 3J7, Canada}

\author[0000-0001-8516-2532]{Tao-Chung Ching}
\affiliation{CAS Key Laboratory of FAST, National Astronomical Observatories, Chinese Academy of Sciences, Peopleʼs Republic of China}
\affiliation{National Astronomical Observatories, Chinese Academy of Sciences, A20 Datun Road, Chaoyang District, Beijing 100012, People's Republic of China} 

\author[0000-0002-0859-0805]{Simon Coud\'{e}}
\affiliation{SOFIA Science Center, Universities Space Research Association, NASA Ames Research Center, Moffett Field, California 94035, USA}

\author[0000-0003-4022-4132]{Woojin Kwon}
\affiliation{Department of Earth Science Education, Seoul National University, 1 Gwanak-ro, Gwanak-gu, Seoul 08826, Republic of Korea}
\affiliation{SNU Astronomy Research Center, Seoul National University, 1 Gwanak-ro, Gwanak-gu, Seoul 08826, Republic of Korea}

\author[0000-0002-6668-974X]{Jia-Wei Wang}
\affiliation{Academia Sinica Institute of Astronomy and Astrophysics, No.1, Sec. 4., Roosevelt Road, Taipei 10617, Taiwan}

\author[0000-0003-1853-0184]{Tetsuo Hasegawa}
\affiliation{National Astronomical Observatory of Japan, National Institutes of Natural Sciences, Osawa, Mitaka, Tokyo 181-8588, Japan}

\author[0000-0001-5522-486X]{Shih-Ping Lai}
\affiliation{Institute of Astronomy and Department of Physics, National Tsing Hua University, Hsinchu 30013, Taiwan}
\affiliation{Academia Sinica Institute of Astronomy and Astrophysics, No.1, Sec. 4., Roosevelt Road, Taipei 10617, Taiwan} 

\author[0000-0002-5093-5088]{Keping Qiu}
\affiliation{School of Astronomy and Space Science, Nanjing University, 163 Xianlin Avenue, Nanjing 210023, Peopleʼs Republic of China}
\affiliation{Key Laboratory of Modern Astronomy and Astrophysics (Nanjing University), Ministry of Education, Nanjing 210023, People's Republic of China}









\begin{abstract}
We present the first $850$\,$\mu$m polarization observations in the 
most active star-forming site of the Rosette Molecular Cloud (RMC, $d\sim1.6$\,kpc)
in the wall of the Rosette Nebula, imaged with the SCUBA-2/POL-2 instruments of 
the JCMT, as part of the B-Fields In Star-Forming Region Observations 2 (BISTRO-2) survey. 
From the POL-2 data we find that the polarization fraction decreases with the $850$\,$\mu$m 
continuum intensity with $\alpha = 0.49 \pm 0.08$ in the $p \propto I^{\rm -\alpha}$ relation,
which suggests that some fraction of the dust grains remain aligned at high densities.
The north of our $850\,\mu$m image reveals a ``gemstone ring'' morphology, which is a  
$\sim 1$\,pc-diameter ring-like structure with extended emission in the ``head'' to the south-west.
We hypothesize that it might have been blown by feedback in its interior, while the B-field is 
parallel to its circumference in most places. 
In the south of our SCUBA-2 field the clumps are apparently connected with filaments which
follow Infrared Dark Clouds (IRDCs).
Here, the POL-2 magnetic field orientations appear bimodal with respect to the large-scale 
{\it Planck} field.
The mass of our effective mapped area is $\sim 174\, M_\odot$ that we calculate from $850$\,$\mu$m 
flux densities. We compare our results with masses from large-scale 
emission-subtracted {\it Herschel} $250$\,$\mu$m data, and find agreement within $30$\%.  
We estimate the POS B-field strength in one typical subregion using the 
Davis-Chandrasekhar-Fermi (DCF) technique and find $80 \pm 30$\,$\mu$G 
toward a clump and its outskirts.
The estimated mass-to-flux ratio of $\lambda = 2.3 \pm 1.0$ suggests that
the B-field is not sufficiently strong to prevent gravitational collapse in this
subregion. 
\end{abstract}



\section{Introduction} \label{sec:intro}

The role of the magnetic field (B-field) through the evolutionary stages of star formation 
from the scale of molecular clouds down to the 
scale of protostars is crucial to our understanding of the star formation process.
In particular, it is not well understood whether 
B-fields help or hinder star formation at each stage and on different spatial scales 
\citep[][and references therein]{Hull+Zhang2019} 

Submillimeter continuum polarization surveys have the potential to allow us to trace dust properties 
and the orientation of the plane-of-sky B-field in molecular clouds
\citep[e.g.,][]{Matthews+2009}. The polarization data, when
complemented with molecular line information, are also a powerful tool 
for estimating the magnetic field 
strength using the Davis-Chandrasekhar-Fermi method \citep{Davis1951, Chandrasekhar+Fermi1953}.   
The JCMT Large Program BISTRO surveys are using the SCUBA-2 bolometer array with its associated 
POL-2 polarimeter, to survey numerous star formation regions \citep{Ward-Thompson+2017}.
The resolution of these surveys ($\sim 14$\arcsec at 850\,$\mu$m) is intermediate between 
the large-scale, low-resolution ($\sim 5$\arcmin) {\it Planck} survey \citep[e.g.,][]{PlanckCollabXXXV+2016} 
and the very high-resolution ($\sim 0.5$\arcsec) small-scale observations of 
interferometers such as ALMA \citep[e.g.,][]{Pattle+2020a}.

The original BISTRO-1 program \citep{Ward-Thompson+2017}
aimed to produce an unbiased survey of the magnetic field in a large sample of typically low-mass
star-forming regions in the solar neighborhood. 
The subsequent BISTRO-2 survey now aims to explore the `mass axis' of star formation parameter 
space by targeting intermediate- and high-mass star-forming regions out to a distance of $\sim 2$\,kpc. 
The BISTRO-1 and 2 programs have generated a homogeneous, statistically significant sample of legacy
observations, with which we are investigating how the 
behaviour of magnetic fields changes from low-mass to high-mass star formation, hence allowing us to 
study the interplay between self-gravity of the gas and other forces. 

\begin{figure*}[!t]
 \begin{center}
 \begin{minipage}{1.0\linewidth}
 \resizebox{1.0\hsize}{!}{\includegraphics[angle=0]{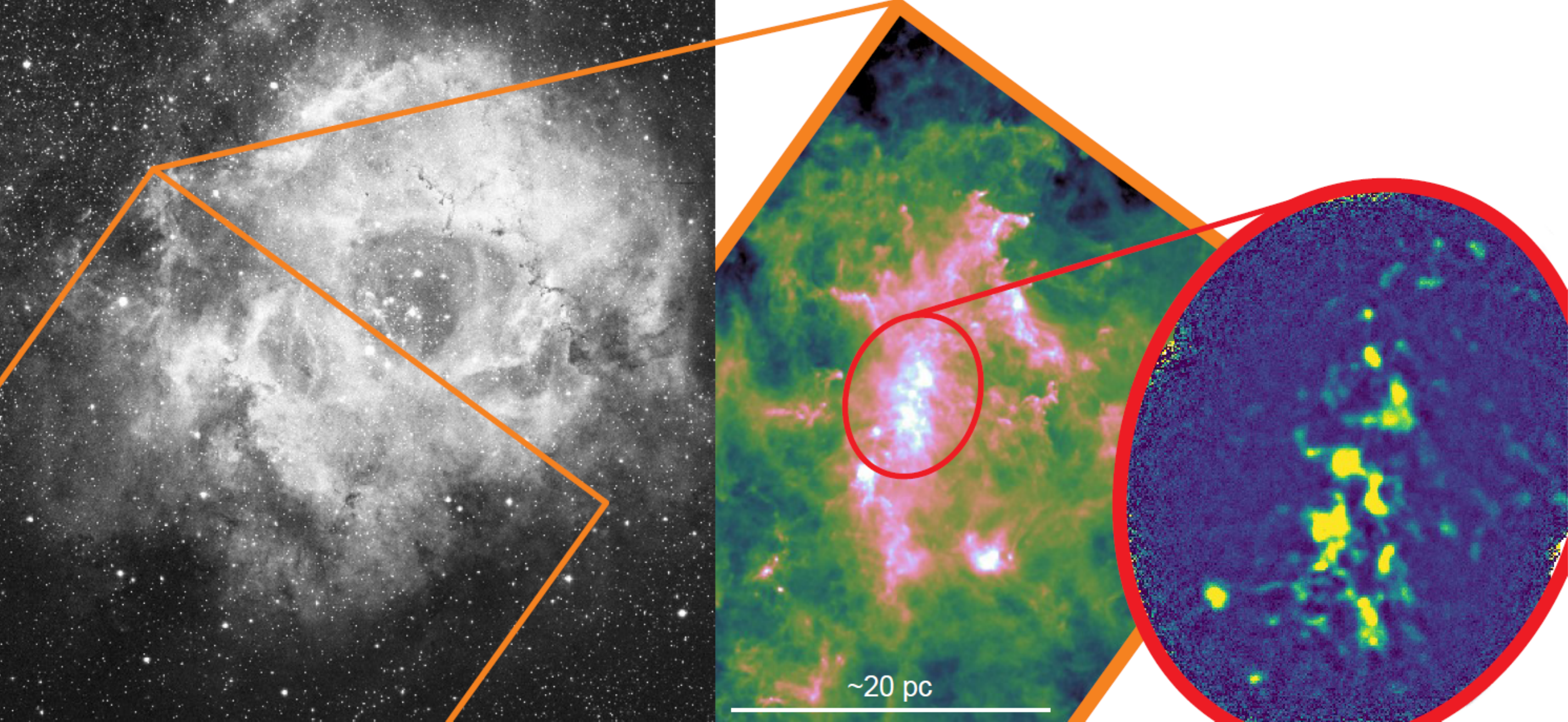}} 
\end{minipage}
 \end{center}
   \caption{The left panel shows the Rosette Nebula in an optical map from the all-sky Digital Sky Survey \citep{Lasker+1990}. 
            The RMC portion of the nebula is shown as a 250\,$\mu$m {\it Herschel} image \citep[e.g.,][]{Schneider+2010} 
            in the middle panel. The zoomed image on the right displays the Stokes $I$ image from our new data of the 
            most active star-forming center of the RMC ($d \sim 1.6$\,kpc) observed by SCUBA-2/POL-2 on JCMT (see Fig.~\ref{fig:Imap}). 
            } 
   \label{fig:bigView}%
\end{figure*}


In this paper we present the first BISTRO-2 results in the high-mass star-forming center of 
the Rosette Molecular Cloud (RMC).
The RMC looks just as a ``petal'' of the Rosette Nebula the cavity of which 
appears to have been blown by the central OB cluster NGC~2244 and its expanding HII-region
(see left panel of Fig.~\ref{fig:bigView}), 
which interacts with the cloud \citep{Roman-Zuniga+Lada2008}. 
This prominent nebula is located in the larger Monoceros OB2 cloud \citep[see][]{Perez1991}, 
in the constellation Monoceros.  

\begin{figure}[!hhhh]
 \begin{center}
 \begin{minipage}{1.\linewidth}
 \resizebox{1.0\hsize}{!}{\includegraphics[angle=0]{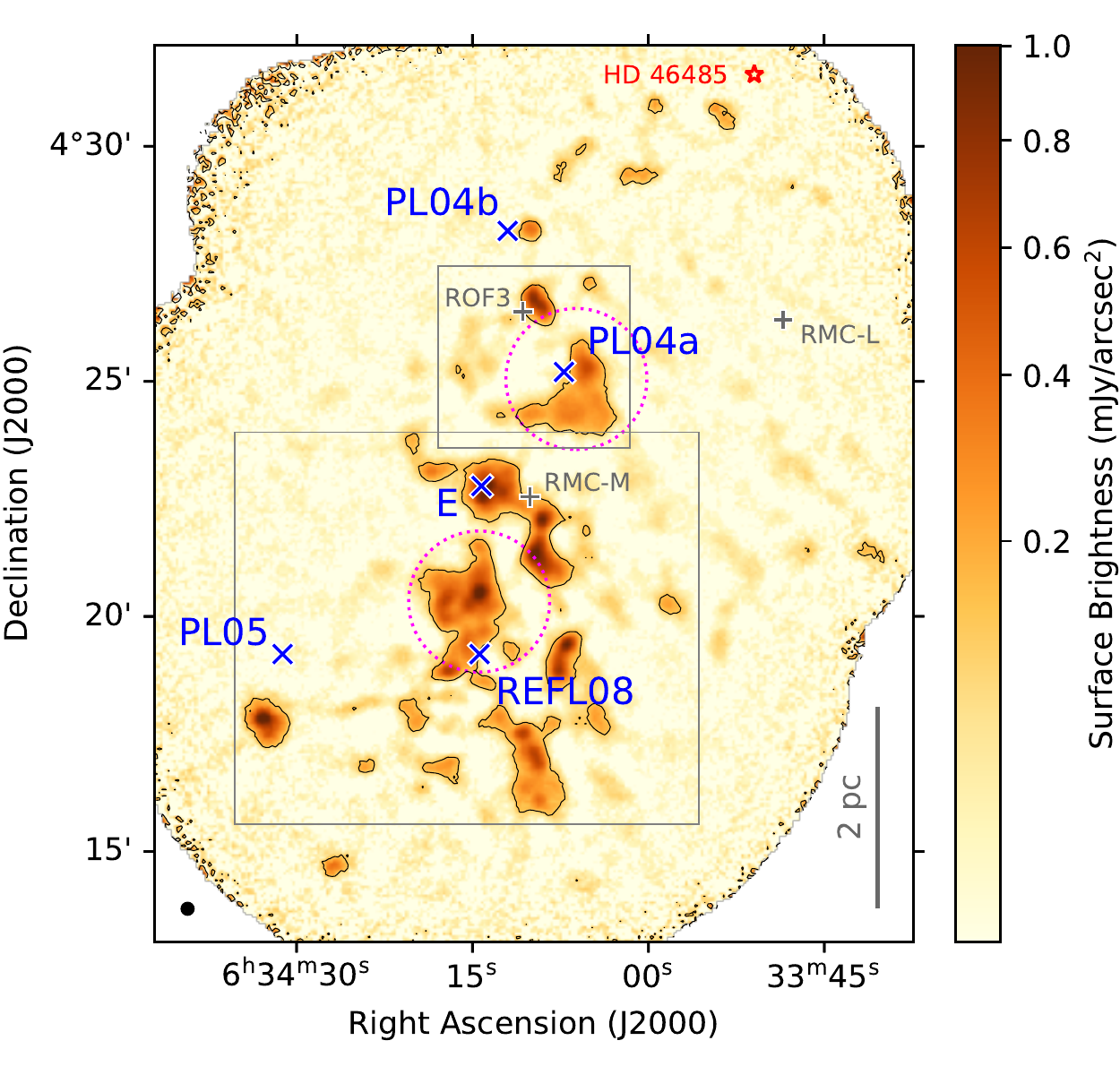}}
\end{minipage}
 \end{center}
   \caption{Stokes $I$ image of the RMC at 850\,$\mu$m, the FWHM resolution of which is $\sim14$\arcsec. 
   The beam is shown in the lower left corner. 
   Dotted magenta circles show the central 3\arcmin-diameter regions, and the black contours
   correspond to $10 \sigma_{I} = 0.13$~mJy/arcsec$^2$.  
   In the following we display results in the outlined boxes.  
   Projected center positions of NIR clusters are marked with blue crosses 
   (PL, REFL, and E indicate cluster positions identified by 
   \citet{Phelps+Lada1997, Roman-Zuniga+2008}, and \citet{Poulton+2008}, respectively).  
   The RMC-M and RMC-L sources marked in gray are possible [SII] outflow features from \citet{Ybarra+Phelps2004}, 
   and ROF3 is a CO outflow feature found by \citet{Dent+2009}.
   The red star shows the position of the closest O-star of NGC~2244 (in projection). 
   } 
   \label{fig:Imap}%
\end{figure}

Distance estimates to NGC~2244 range from $1.4$ to $1.7$\,kpc \citep{Ogura+Ishida1981, Perez+1987,
Hensberge+2000, Park+Sung2002, Lombardi+2011, Martins+2012, Bell+2013, Kharchenko+2013}. 
Gaia DR2 yields a distance estimate to NGC~2244 of $1.59$\,kpc with
a $1$\% statistical error and an 11\% systematic error \citep{Muzic+2019}. 
In this work, therefore, we adopt $\sim 1.6$\,kpc as the distance of the RMC.
The stellar content of the cluster has been extensively studied from X-ray to mid-infrared wavelengths.
Seven main O-stars are thought to be evacuating the central part of the nebula as part of the 
ionizing cluster NGC~2244 \citep{Martins+2012}, 
the closest of which (only in projection), HD~46485, is marked in Fig.~\ref{fig:Imap}. 
The cluster's young age, $\sim 2$\,Myr \citep[e.g.,][]{Park+Sung2002, Bell+2013}, 
together with the absence of non-thermal 
radio emission, lead to the conclusion that no supernova explosion has occurred 
yet in the nebula \citep{Townsley+2003}.

Many embedded clusters have been identified in the RMC. Seven of them (PL01-07) were found
by \citet{Phelps+Lada1997} at near-infrared (NIR) 
wavelengths. \citet{Roman-Zuniga+2008} discovered two more NIR clusters, REFL08 and REFL09.  
\citet{Poulton+2008} defined clusters with IR excess sources from A to G, 
extending from a few to several square parsecs. 
The center of the RMC, that we observed with JCMT, contains most of the cluster members of 
PL04a/b, PL05, and REFL08, which are associated with a single CO clump identified 
by \citet{Williams+1995}. The cluster E identified by \citet{Poulton+2008} covers 
most of our observed region except PL04a/b and PL05 (see their Figure~10.). 
In PL04a (see Fig.~\ref{fig:Imap}) near-infrared-excess sources coincide spatially with 
the NIR nebulosity \citep{Roman-Zuniga+2008}. 
Around PL04b, X-ray sources have been found with $Chandra$ by \citet{Wang+2009}, 
which indicate the presence of Class III young stellar objects (YSOs).

\citet{Roman-Zuniga+2008} estimated that the gas-rich clusters of the 
RMC center provide half of the star formation in the 
whole cloud. This subregion is beyond the ionization front of the HII region where 
a shock front may have already passed through.
From their NIR survey they found that the age of the cluster members 
decreases with increasing distance from the Rosette Nebula. 
\citet{Roman-Zuniga+2008} hypothesised that
the origin of the age sequence with small age differences is primordial, a result 
of the formation and evolution of the cloud, and not of the HII region. 
This result has been confirmed by \citet{Poulton+2008}, \citet{Ybarra+2013}, and \citet{Cambresy+2013}.
Based on near-infrared $JHK_{\rm S}$ and WISE data, \citet{Cambresy+2013} also find that the age distribution 
of the young clusters in the region is not consistent with a triggered star formation scenario, and they 
conclude that the evolution of the Rosette complex is not governed by the influence of the central OB star 
population.  
Interestingly, the cloud collapse may have been triggered externally which then formed 
the dense ridge, located along the mid-plane of the cloud (see the middle panel 
of Fig.~\ref{fig:bigView}), and ignited star formation \citep{Poulton+2008}.

Far-infrared and submillimeter {\it Herschel} HOBYS data \citep{Motte+2010} of the Rosette 
region also shed light on the influence of NGC~2244 on the cloud \citep{Schneider+2010, Tremblin+2013, Tremblin+2014}. 
The authors present the properties of embedded protostellar sources \citep{Hennemann+2010}, 
and assess the clump populations up to $1$\,pc in size \citep{DiFrancesco+2010}. 
From the distribution of starless and protostellar clumps, the latter 
authors did not find an age-gradient across the RMC. 
However, \citet{Schneider+2010} tentatively conclude from the spatial distribution 
of the most massive dense cores ($0.05-0.3$\,pc) that there may be an age sequence 
with younger cores further away from NGC~2244 
that is consistent with the above findings of \citet{Poulton+2008}, \citet{Roman-Zuniga+2008},
\citet{Ybarra+2013}, and \citet{Cambresy+2013}.  
\citet{Schneider+2012} investigate the filamentary structure of the RMC and propose that the 
sites of star-cluster formation correlate with the junctions of the filamentary network. 
For part of the {\it Herschel} coverage in Rosette, see the middle panel of Fig.~\ref{fig:bigView}.

\citet{PlanckCollabXXXIV+2016} traced the 3D magnetic field structure of Rosette with 
{\it Planck} polarization data combined 
with rotation measure (RM) observations from \citet{Savage+2013} to trace the magnetic 
field at low resolution both in the molecular and ionized parts of the cloud. 
The analytical model of \citet{PlanckCollabXXXIV+2016}
reproduced the large-scale mean observed properties in the Rosette, 
such as the rotation measure distribution and mean dust polarization values.
 
These {\it Planck} observations show that the large-scale
magnetic field in the Rosette's parent molecular cloud is mostly parallel to the large-scale 
field along the Galactic plane. 
\citet{PlanckCollabXXXIV+2016} found overall low polarization fractions 
in and around the Rosette Nebula, typically $p<6$\%,
with the lowest values ($p\lesssim 3$\%) toward the densest regions. 
They estimate a line-of-sight (LOS) B-field strength of $\sim 3\,\mu$G from rotation measure data.
The strength and structure of the magnetic field in 
Rosette was also estimated 
by \citet{Costa+2016} from Faraday rotation measurements of extragalactic radio sources through 
the nebula. In agreement with earlier results, they also detect an excess rotation measure at 
the shell of the Rosette Nebula.

\begin{figure*}[!t]
 \begin{center}
 \begin{minipage}{1.\linewidth}
 \hspace{-4mm}
 \resizebox{0.515\hsize}{!}{\includegraphics[angle=0]{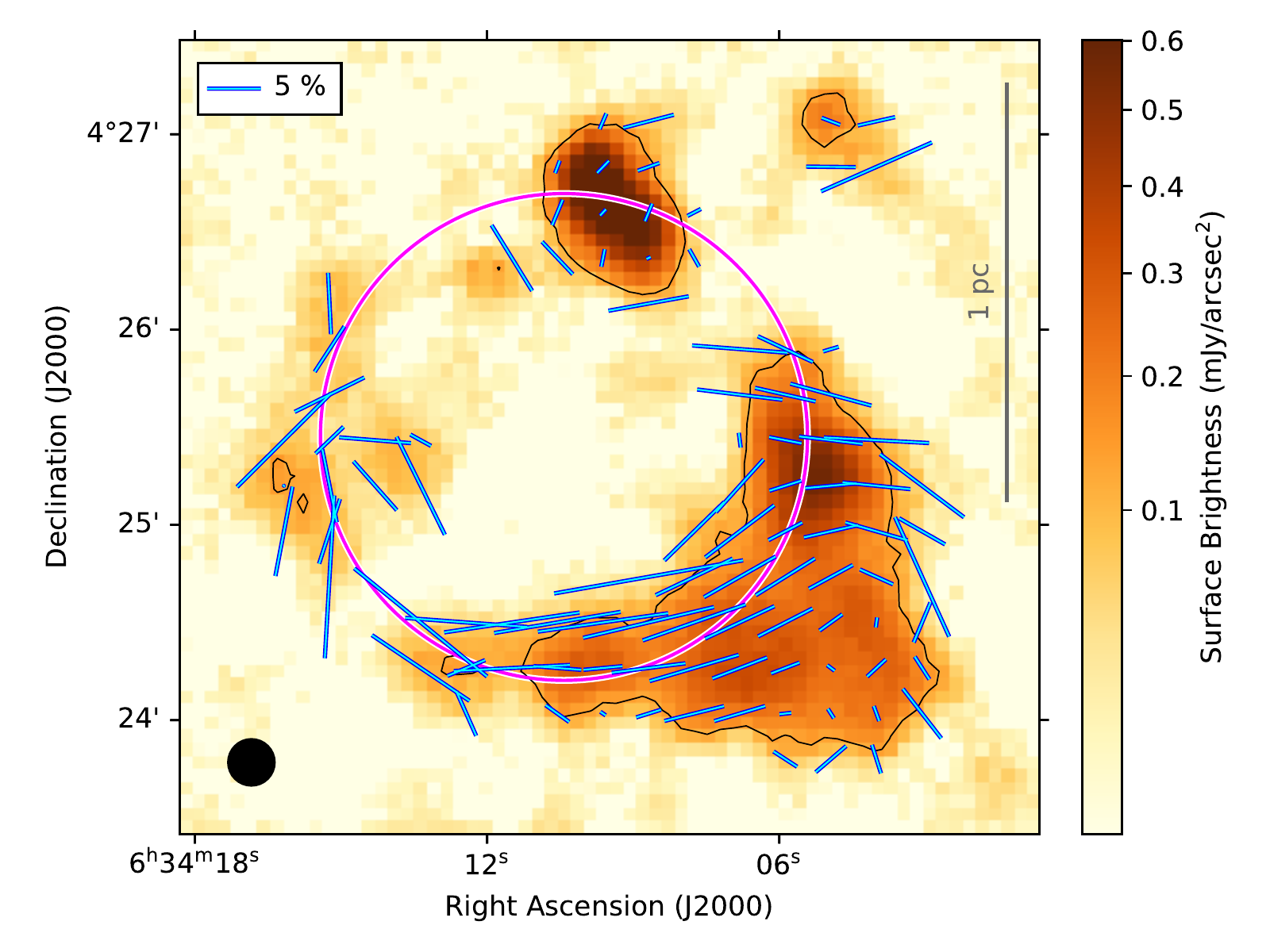}} 
 \hspace{-2mm}
 \resizebox{0.515\hsize}{!}{\includegraphics[angle=0]{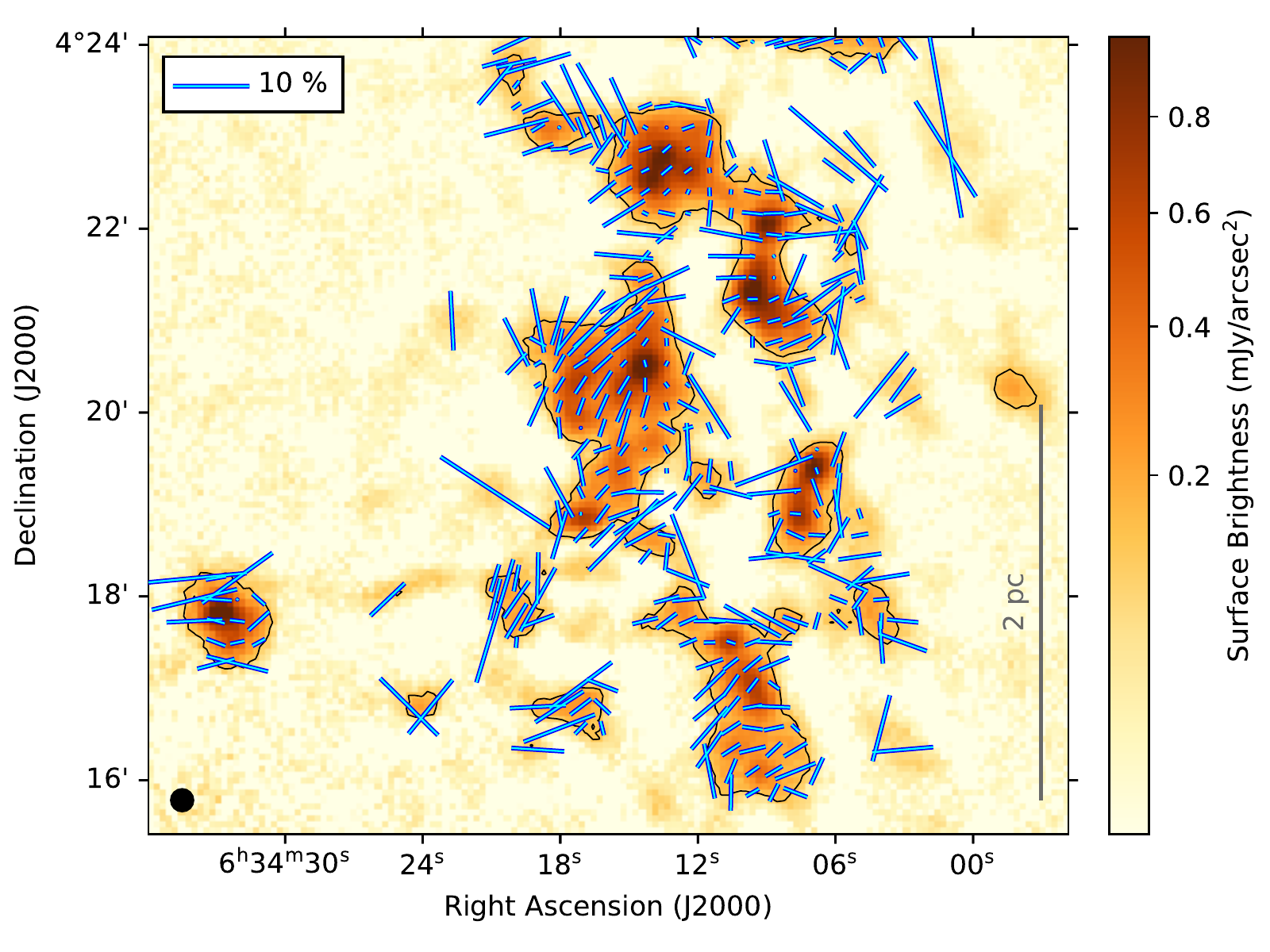}}
\end{minipage}
 \end{center}
   \caption{Maps of debiased polarization half-vectors in the central part of the RMC,
   coarsely selected with the criteria of $I > 0$, $I/\delta I > 10$, and $\delta p < 5\%$.
   The lengths of the POL-2 half-vectors in blue are proportional to their polarization fractions, the scale
   of which is shown in the map panels. 
   The background is a SCUBA-2 $850$\,$\mu$m Stokes $I$ image, where the black contours are as in Fig.~\ref{fig:Imap}.
   {\bf Left:} Polarization map of the northern field, featuring the ring-like structure that is indicated by 
   the magenta circle (see Fig.~\ref{fig:Imap}). 
   {\bf Right:} Polarization map of the southern field (also see Fig.~\ref{fig:Imap}).
   }  
   \label{fig:pol14as_IdI10}%
\end{figure*}

\begin{figure*}[!t]
 \begin{center}
 \begin{minipage}{1.\linewidth}
 \hspace{-4mm}
 \resizebox{0.515\hsize}{!}{\includegraphics[angle=0]{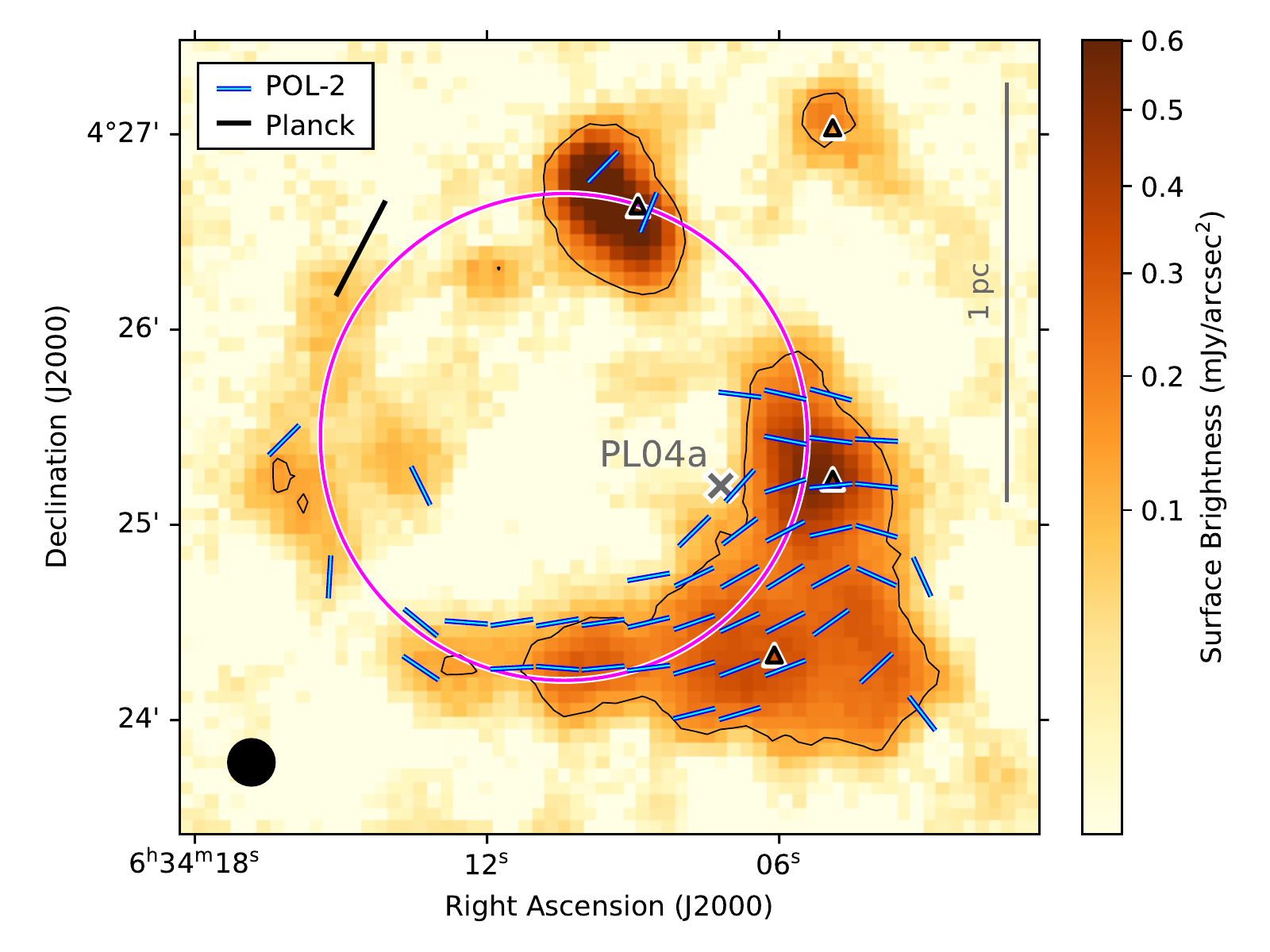}} 
 \hspace{-2mm}
 \resizebox{0.515\hsize}{!}{\includegraphics[angle=0]{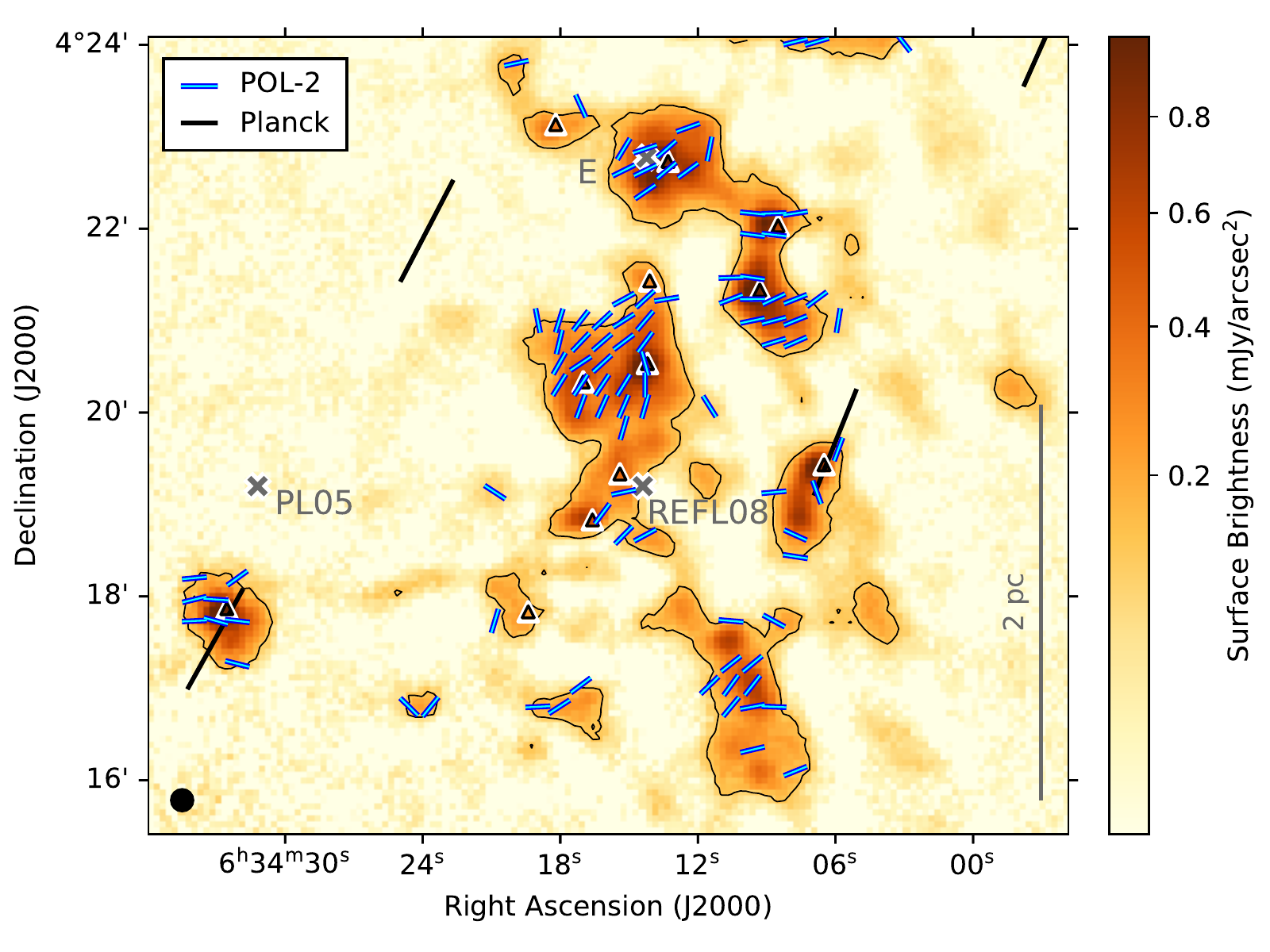}}
\end{minipage}
 \end{center}
   \caption{Maps of polarization half-vectors, rotated by 90\degr~to show the orientation of the B-field,
   selected with the criteria of $I > 0$, $I/\delta I > 10$, $p/\delta p > 3$, and $\delta p < 5\%$ (see
   Sect.~\ref{sec:polprop}). 
   The POL-2 magnetic half-vectors in blue have equal lengths in each panel, and are plotted on a 
   14\arcsec~vector grid. 
   {\it Planck} B-field-oriented vectors (on a 5\arcmin-scale) are shown in black, which are 
   approximately parallel to the Galactic plane, as found by \citet{PlanckCollabXXXIV+2016}.
   The background is a SCUBA-2 $850$\,$\mu$m Stokes $I$ image, where the black contours are as in Fig.~\ref{fig:Imap}.
   Triangle symbols mark the submillimeter-continuum objects detected by SCUBA \citep{DiFrancesco+2008}.
   {\bf Left:} Magnetic-field vector map of the northern field, featuring the ring-like structure of 
   PL04a, indicated by the magenta circle (see Fig.~\ref{fig:Imap} and text). 
   {\bf Right:} Magnetic-field vector map of the southern field (also see Fig.~\ref{fig:Imap}).
   } 
   \label{fig:pol14as}%
\end{figure*}

We present here the first results from the BISTRO-2 survey of the actively star-forming RMC center.
The paper is organized as follows. 
Section~\ref{sec:obs} provides details about the JCMT observations and the data reduction.
In Section~\ref{sec:res} we present the polarization properties, and the magnetic field morphology.
In Section~\ref{sec:discuss} we derive and discuss the mass of the region, and the 
B-field strength with DCF analysis.
Finally, Section~\ref{sec:conc} presents our main conclusions.

\section{Observations and Data Reduction} \label{sec:obs}

As part of the JCMT BISTRO-2 survey, the central part of the Rosette molecular cloud was observed at 850\,$\mu$m with SCUBA-2 
\citep{Holland+2013} and POL-2 \citep{Friberg+2016} between 12 January 2019 and 02 May 2019, under 
JCMT project code M17BL011.
The region was observed in two overlapping tiles, each was observed
20 times for $\sim 40$ minutes each time,
giving a total on-source integration time of $\sim 27$ hours.
The two overlapping observations were made with the POL-2 DAISY mode \citep{Friberg+2016}, 
which produces a map with high signal-to-noise ratio (SNR) in the 
central 3\arcmin-diameter region with increasing noise to the edges. 
These were combined during the data reduction with the `multi-object' keyword on.
Bad datasets were not found among the observations.
During the observations the atmospheric opacity, $\tau$ at 225\,GHz, varied 
between $\sim 0.02$ and $\sim 0.07$.

The effective beam size of JCMT is $14.1$\arcsec~($\sim 0.1$\,pc at $1.6$\,kpc) at 850\,$\mu$m. 
Continuum polarimetric observations were 
simultaneously taken at $450$\,$\mu$m with a resolution of $9.6$\arcsec, although those data will be
presented in a future publication; and in this paper we only discuss the $850$\,$\mu$m dataset. 

The $850$\,$\mu$m data were reduced using the SMURF \citep{Jenness+2013, Berry+2005} package in 
Starlink \citep{Currie+2014}. 
In short, the $calcqu$ command of the SMURF package was used first to convert the raw bolometer data 
into Stokes $I$, $Q$, and $U$ time-streams. 
Then, all the time-streams of the observations were co-added into a first solution
Stokes $I$ map with the $makemap$ routine 
inside the $pol2map$ script of the SMURF package. 
Re-running this task creates the final improved $I$ map from the first $I$ map 
solution \citep{Jenness+2013, Berry+2005}.
Finally, $makemap$ is also used for creating the $Q$ and $U$ maps from 
their time-streams, along with their variance maps, and the 
polarization half-vector catalog \citep{Mairs+2015, Pattle+2017}. 
The term `half-vector' is used because of the $\pm 180$\degr~ambiguity 
in the inferred magnetic field direction \citep[e.g.,][]{Kirk+2006, Pattle+2017}
-- i.e., we do not know which end of the half-vector to put the `arrow' on.
The final improved Stokes $I$ map, adopting the ``January 2018'' instrumental polarization model \citep{Friberg+2018}, 
was used to help correct for the instrumental polarization in the $Q$ and $U$ maps.

The final Stokes $I$, $Q$, $U$ maps and the polarization catalog are gridded to a default 4\arcsec/pixel scale. 
In these maps we estimated the one-sigma sensitivities $\sigma_{I}$, $\sigma_{Q}$, $\sigma_{U}$ 
to be 2.9, 2.4, and 2.2~mJy/beam, respectively.
The corresponding uncertainties in the respective order are 0.013, 0.011, and 0.010 in mJy/arcsec$^2$, 
and the level of $10 \sigma_{I} = 0.13$~mJy/arcsec$^2$ is marked in our Stokes $I$ figures.

With the $pol2map$ binsize parameter we generated a catalog of independent polarization vectors binned to a 14\arcsec/pixel 
scale (to match the beam size), 
while for the Stokes $I$ map we use at the default 4\arcsec/pixel scale (to produce a smoother-looking image). 
The data acquisition and reduction, as well as the absolute calibration of the data are discussed in detail by \citet{Ward-Thompson+2017}.

\section{Results and Analysis} \label{sec:res}

\subsection{Polarization properties} \label{sec:polprop}

The BISTRO-2 observations with SCUBA-2 and POL-2 cover the most active star-forming site 
in the wall of the Rosette Nebula within 
an effective area of $\sim 0.06$ square degrees, or $\sim 45$\,pc$^2$ at a distance of $\sim 1.6$\,kpc.
See Fig.~\ref{fig:bigView} for a large view of the Rosette Nebula, and Fig.~\ref{fig:Imap} 
for the $850$\,$\mu$m Stokes $I$ map toward the center of the RMC made with SCUBA-2. 

We follow the conventional definitions of the polarization properties \citep[e.g.,][]{Pattle+2017, Coude+2019}. 
The measured polarization angles are defined as $\theta$ $=$ $0.5~{\rm arctan}\,(U/Q)$.
The non-debiased polarized intensity is $I_{\rm p}$ = $(Q^2+U^2)^{0.5}$, and the corresponding polarization 
fraction is defined as $p$ = $I_{\rm p} / I$.
The debiased polarized intensity, however, is calculated as $I^{\rm db}_{\rm p}$ = $(Q^2+U^2-0.5[(\delta Q)^2+(\delta U)^2])^{0.5}$,
with $\delta Q = \sqrt(V_Q)$ and $\delta U = \sqrt(U_Q)$, where $V_Q$ and $U_Q$ are the variances of $Q$ and $U$.
The debiased polarization fraction is then given as $p^{\rm db}$ = $I^{\rm db}_{\rm p} / I$. 

In Fig.~\ref{fig:pol14as_IdI10}, we show a more complete set of our polarization data that we 
coarsely selected with the criteria of Stokes $I > 0$, $I/\delta I > 10$, and $\delta p < 5\%$.
Here, the debiased polarization half-vectors also preserve the information on the percentage polarization.   
The polarization vector field seems ordered in the higher Stokes $I$ -- and a priori denser -- regions, 
and the polarization fraction appears to decrease with increasing density (see Sect.~\ref{sec:pI_rel}). 

However, for most of the following analysis, we use the vector selection criteria of
Stokes $I > 0$, $I/\delta I > 10$, $p^{\rm db}/\delta p > 3$, and $\delta p < 5\%$,
where $\delta I$ and $\delta p$ indicate the uncertainty in total intensity and 
polarization fraction (both non-debiased and debiased), respectively. 
This set of independent criteria, giving us 152 vectors at 14\arcsec~binning, were adapted 
from the criteria used in, for example, \citet{Coude+2019}.  

We consider polarization half-vectors rotated by 90\degr~to trace the magnetic field direction 
that we refer to as ``magnetic field half-vectors'' in the plane of the sky. 
This can be assumed, however, only if the dust grain size is much smaller than the observed 
wavelength \citep{Kirchschlager+2019, Guillet+2020}. Then, the emitting elongated dust grains are 
mostly aligned by the magnetic field, and the magnetic field direction is orthogonal to the 
polarization direction \citep[e.g.,][]{Lazarian&Hoang2007, Hoang&Lazarian2016}.

The above-selected POL-2 magnetic half-vectors in the plane-of-sky (POS) are shown in
Fig.~\ref{fig:pol14as}, overlaid on our Stokes $I$ map. 

\subsection{$p-I$ relationship} \label{sec:pI_rel}

Dust grain alignment efficiency can be assessed using the relationship 
between polarization efficiency and visual extinction \citep[e.g.,][]{Whittet+2008, Jones+2015}.  
For optically thin submillimeter emission polarimetry, this is commonly 
treated as a relationship between the polarization fraction and total intensity
(e.g., \citealt{Jones+2015}).   
Observations of polarized dust emission typically show a power-law dependence, 
$p\propto I^{-\alpha}$, where $0\leq \alpha \leq 1$.  
A steeper index (higher $\alpha$) indicates poorer grain alignment; 
$\alpha = 0$ indicates that grains are equally well aligned at all depths, 
while $\alpha = 1$ indicates either a lack of aligned grains, or that all 
observed polarized emission is produced in a thin layer at the cloud's 
surface \citep[][and references therein]{Pattle+2019a}.

In order to avoid selection biases which may influence these 
relationships, we infer dust grain alignment properties from Ricean statistics.
We measured $\alpha$ using the method described by \citet{Pattle+2019a}, 
in which we assume that the underlying relationship between \emph{non-debiased} 
$p$ and $I$ can be parameterized as
\begin{equation}
  p = p_{\sigma_{QU}}\left(\frac{I}{\sigma_{QU}}\right)^{-\alpha}
  \label{eq:polfrac}
\end{equation}
where $p_{\sigma_{QU}}$ is the polarization fraction at the RMS noise 
level of the data $\sigma_{QU}$, and $\alpha$ is a power-law index in 
the range $0 \leq \alpha \leq 1$. We fitted the relationship between 
$I$ and observed polarization fraction $p^{\prime}$ with the mean of 
the Ricean distribution of observed values of $p$ which would arise 
from Eq.~\ref{eq:polfrac} in the presence of Gaussian RMS noise 
$\sigma_{QU}$ in Stokes $Q$ and $U$:
\begin{equation}
  p^{\prime}(I) = \sqrt{\frac{\pi}{2}}\left(\frac{I}{\sigma_{QU}}\right)^{-1}\mathcal{L}_{\frac{1}{2}}\left(-\frac{p_{\sigma_{QU}}^{2}}{2}\left(\frac{I}{\sigma_{QU}}\right)^{2(1-\alpha)}\right),
  \label{eq:rmfit}
\end{equation}
where $\mathcal{L}_{\frac{1}{2}}$ is a Laguerre polynomial of order 
$\frac{1}{2}$. See \citet{Pattle+2019a} for a derivation of this result.  
We restricted our dataset to the central 3-arcminute diameter region 
around each pointing centre over which exposure time, and so RMS noise, 
is approximately constant \citep{Friberg+2016}.  
We estimated an RMS noise value in our Stokes $Q$ and $U$ data of 
$0.62\,$mJy/beam on 12\arcsec~pixels, and $p_{\sigma_{QU}} = 0.36\pm 0.14$
for polarization fraction at this noise level.

\begin{figure}[!ttt]
 \begin{center}
 \begin{minipage}{1.\linewidth}
 \resizebox{1.0\hsize}{!}{\includegraphics[angle=0]{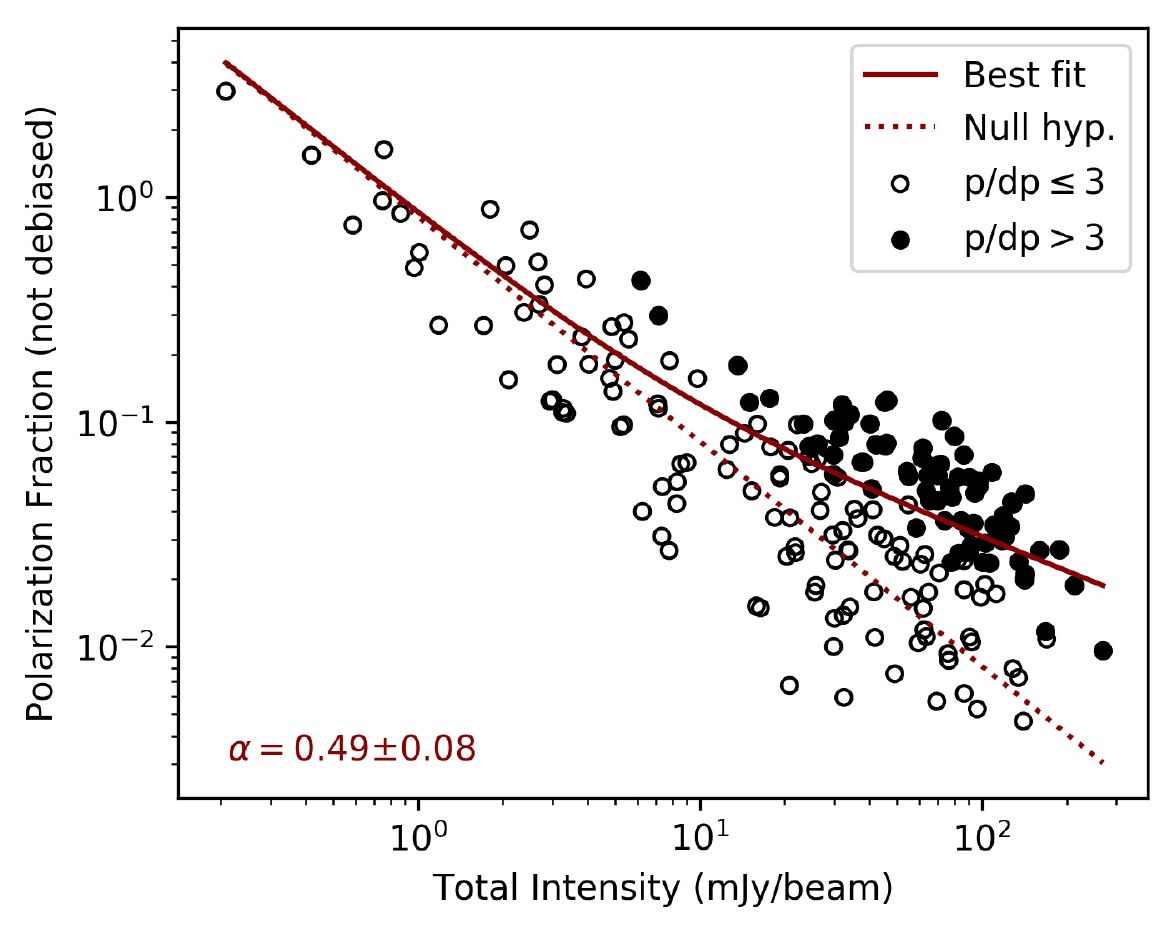}} 
\end{minipage}
 \end{center}
   \caption{Non-debiased polarization fraction $p$ as a function of total intensity at $850\,\mu$m
   fitted with the mean of the Ricean distribution of $p$. 
   All of the data points above $I = 0$ were fitted within the central 3\arcmin-diameter regions of the 
   combined map (see Fig.~\ref{fig:Imap}), the $p/\delta p > 3$ points (filled circles) are marked for 
   information.
   The red solid line gives the best-fit model with $\alpha = 0.49\pm 0.08$, and the dashed line shows
   the null hypothesis; the expected behavior of non-aligned dust grains. 
           } 
   \label{fig:PvsI_14as}%
\end{figure}

Figure~\ref{fig:PvsI_14as} shows the $p-I$ relationship in 
the central ridge of the RMC, and for the central regions of our
observed field.
We measure a best-fit index of $\alpha = 0.49\pm 0.08$, i.e, 
$p$ $\propto$ $I^{-0.49\pm0.08}$.
This suggests that in the RMC, dust grain alignment efficiency decreases 
approximately linearly with increasing density (cf., \citealt{Jones+2015}),
but that some fraction of the grains remain aligned with respect to the 
magnetic field to highest densities.
The partially aligned nature of the dust grains at high densities 
is also supported by the strongly correlated position angles of the
polarization half-vectors which 
we observe (see Fig.~\ref{fig:pol14as_IdI10}).
%

\subsection{Magnetic Field Morphology} \label{sec:morph}

We obtained the POS magnetic field half-vectors by rotating the polarization 
half-vectors by 90\degr.
The magnetic field orientations with equal length vectors are shown 
in Fig.~\ref{fig:pol14as} in the northern and southern map portions which cover 
most of the $850$\,$\mu$m emission in the central ridge of the RMC.

In the following, based on Stokes $I$ and associated data, we describe 
a ring-like structure in the northern part of the observed field, and the system 
of clumps and elongated features in the observed south. We refer to their positions 
mainly with respect to the projected centers of NIR clusters listed in 
Sect.~\ref{sec:intro}.   

\subsubsection{A ring seen by SCUBA-2} \label{sec:ring}

In the north of the region, around PL04a, the $850$\,$\mu$m emission reveals a 
ring-like structure with a diameter of $\sim 1$\,pc.
It is traced by a dense clump in the north of the ring, weaker emission in the east,
and strong clumpy emission extending away from the ring in the south-west. 
This latter corner looks just as a ``gemstone head of a ring with side stones''
(see the left panel of Fig.~\ref{fig:pol14as}). 


The B-field seems to trace the circumference of the ring in the south and weakly in 
the east (where we have sufficient signal-to-noise ratio to plot vectors.
However, see Fig.~\ref{fig:pol14as_IdI10}/left for more polarization vectors 
along the ring). 
In the western part of the ring, where there is a slightly brighter clump, the B-field
appears to run almost perpendicular to the circumference; a similar 
pattern is seen
in the north, but there it is based on 
low numbers of half-vectors. 
The pattern around the ``gemstone'' head is less clear because, again, we appear to have 
insufficient signal-to-noise ratio to plot sufficient number of half-vectors.

\begin{figure}[!th]
 \begin{center}
 \begin{minipage}{1.0\linewidth} 
 \resizebox{1.\hsize}{!}{\includegraphics[angle=0]{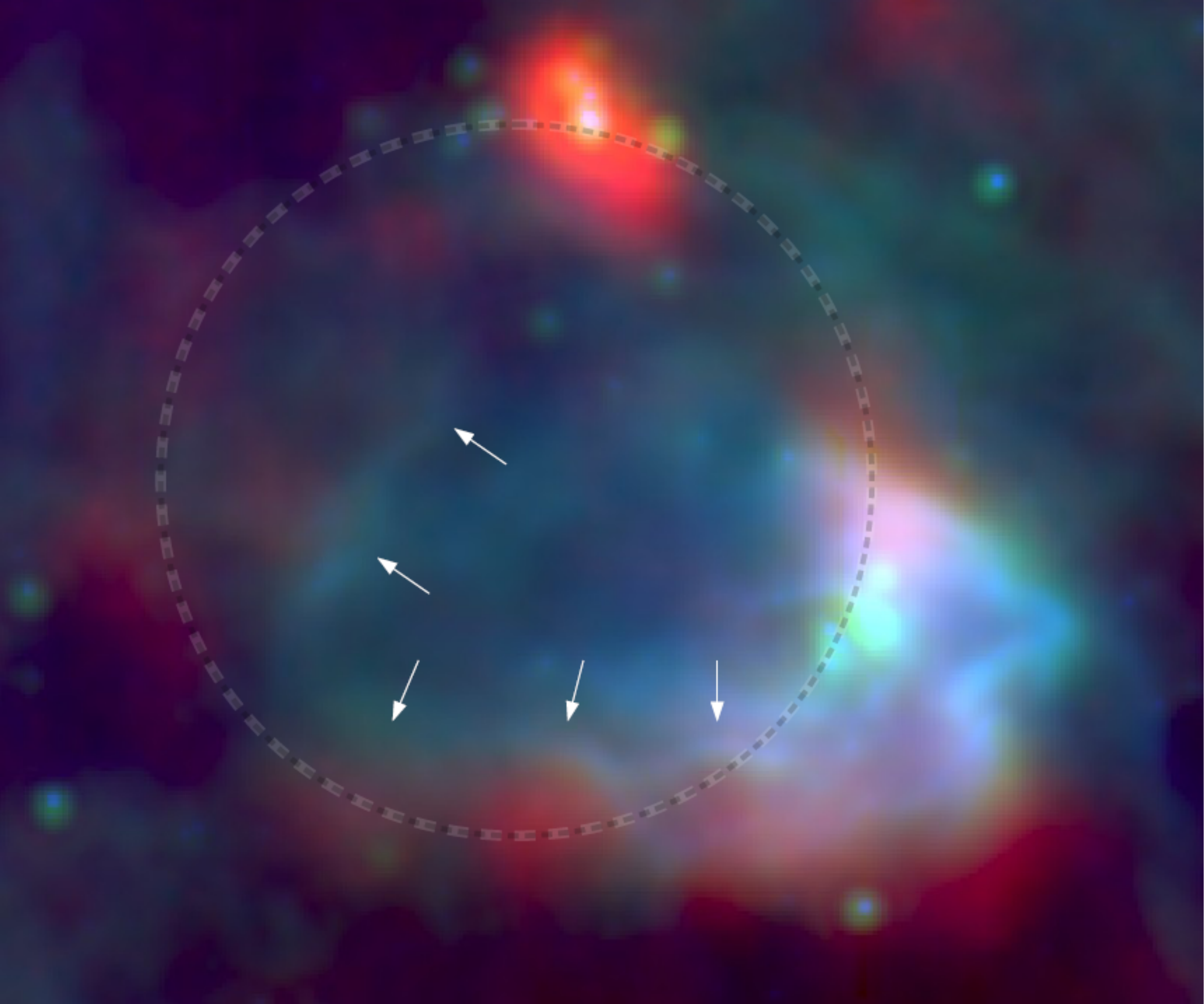}}
\end{minipage}
 \end{center}
   \caption{
            RGB composite image of the SCUBA-2 ``gemstone ring'', where $850$\,$\mu$m
            emission of cold dust is red, $24$\,$\mu$m hot dust emission is green, and 
            the $8$\,$\mu$m PAH emission is blue.
            The overplotted ring is the same as in Fig.~\ref{fig:pol14as}/left,
            and the arrows show where the short-wavelength emission seems to interact
            with the $850$\,$\mu$m emission of the dense clumps (see text for details). 
           } 
   \label{fig:ring}%
\end{figure}

This ring morphology that we see in our Stokes $I$ image, 
is also visible at shorter wavelengths. 
The {\it Spitzer} IRAC/MIPS data from $3.6$\,$\mu$m to $24$\,$\mu$m reveal
emission around a cluster of stars at the western/south-western position 
along the ring (see the bright cyan sources in Fig.~\ref{fig:ring}, and 
the {\it Spitzer}-only image in Figure~2 of \citet{Poulton+2008}).
At these bright sources, 2MASS \citep{Cutri+2003}, WISE \citep{Cutri+2012, Cambresy+2013},
and other IR \citep{Bica+2003, Phelps+Lada1997} star clusters are registered,
with (candidate) YSOs around. 
For the distribution of YSOs in our whole observed field,
see Fig~\ref{fig:8mu}.

The short-wavelength emission of the YSO cluster seems to illuminate 
and fill the interior of our ring. 
We have marked with arrows in Fig.~\ref{fig:ring} where it appears that 
the {\it Spitzer} emission ($8$\,$\mu$m -- blue, $24$\,$\mu$m -- green)
is in direct interaction with the $850$\,$\mu$m (red) clumps. The arc-shaped 
red-green-blue gradients along the arrows probably show us the penetration
of the short-wavelength emission from the illuminating cluster sources
(in the south-west of the ring) into the dense cloud material. 

We speculate that this process might have shaped the $850$\,$\mu$m dense 
material not only at the ``gemstone'' and ``side stones'', but in a large
part of the ring.
The short-wavelength bow-shock-shaped emission 
(in cyan in Fig.~\ref{fig:ring}) may be due to a break out of the clumpy 
ring/bubble toward the west and toward the observer 
(see Fig.~\ref{fig:pol14as}/left and Fig.~\ref{fig:ring}). 
However, this hypothesis needs to be further investigated.  

HII regions can be identified using mid-infrared (MIR) wavelengths as well.
Galactic HII regions are typically characterized by a rim-like $\sim 10\,\micron$
emission surrounding bubble-like radiation at $\sim 20\,\micron$ that coincide 
with the ionized gas. 
\citep[e.g.,][]{Povich+2007, Anderson+2011, Simpson+2012}.

The $\sim 10-20\,\micron$ emission is from polycyclic aromatic hydrocarbon (PAH) 
molecules which fluoresce in the presence of ultraviolet radiation fields, and
can thus be identified in $8$\,$\mu$m and $24$\,$\mu$m images. 
Fig.~\ref{fig:ring} shows similar ``layered'' MIR features in an $850$\,$\mu$m ring-like
structure, and Fig~\ref{fig:8mu} shows the distribution of the $8$\,$\mu$m emission
on a larger scale.  

The ring itself looks like a cavity blown by feedback in its interior, and the B-field is 
parallel to the circumference of the arc in most places.
Similar ``curved'' magnetic field geometry was found in the ring-like shell of bubble N4 by 
\citet{Chen+2017} from NIR polarization.
In their Radiation MHD simulations of HII regions, \citet{Arthur+2011} also witness mostly 
parallel orientations of the magnetic fields to the shell and ionisation front.

We also note that the H$_{\rm 2}$ column density values (Ladjelate et al., in prep.)
in the whole observed region are everywhere above the inferred threshold of 
log$_{\rm 10}$($N_{\rm H}$) $\approx$ 21.7\,cm$^{-2}$ (or $\sim 2.5 \times 10^{21}$\,cm$^{-2}$ 
in $N_{\rm H2}$) where the preferred relative orientation between the B-field and density 
structures change from parallel to perpendicular \citep{PlanckCollabXXXV+2016}.
Therefore, neither in the ring, nor in the observed field can we test this {\it Planck} threshold.

\subsubsection{The southern field} \label{sec:south}

In the southern part of the RMC center, Dec(2000) $\lesssim$ 4\degr 24\arcmin~in Fig.~\ref{fig:Imap},
the SCUBA-2 field looks clumpy. This can also be seen in the right panel of Fig.~\ref{fig:pol14as}, 
overplotted with the submillimeter-continuum objects detected by SCUBA \citep{DiFrancesco+2008}.
Some of the SCUBA/SCUBA-2 clumps seem to be connected with each other by 850-$\mu$m emission 
filamentary features.

Here, the selected POL-2 magnetic half-vectors appear to be ordered at higher $850$\,$\mu$m 
emission and they roughly follow the {\it Planck} B-field in the following areas: 
at the center position of 
cluster E; between cluster E and REFL08; and south-west of REFL08. 
Our B-field seems to turn roughly east-west in the other 
two clumps (south of PL05, and south-west of cluster E).
So, the field geometry appears bimodal with some clumps well aligned with the large-scale field 
and some clumps nearly perpendicular. 

\begin{figure}[!th]
 \begin{center}
 \begin{minipage}{1.\linewidth}
 \resizebox{1.0\hsize}{!}{\includegraphics[angle=0]{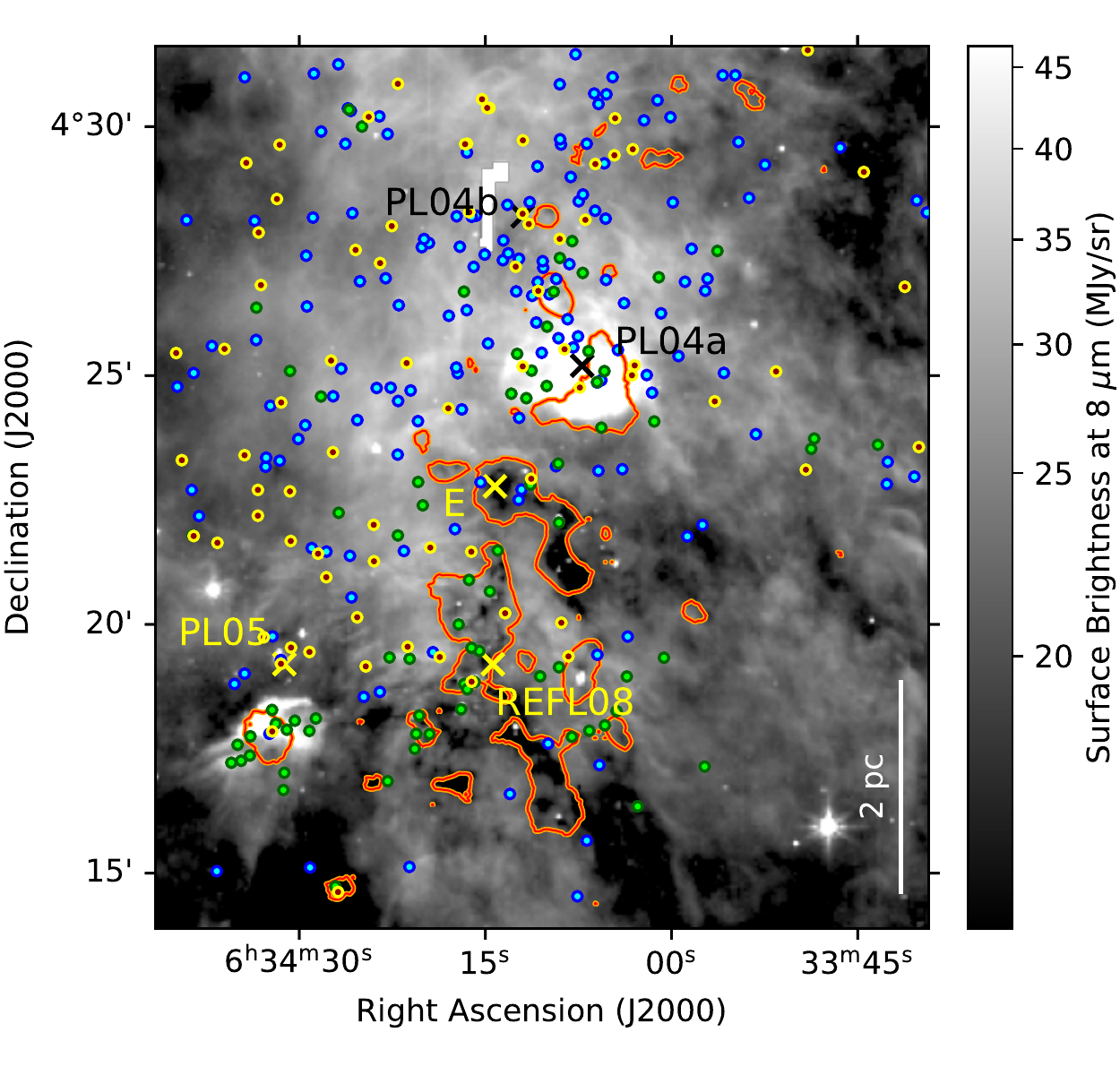}}
\end{minipage}
 \end{center}
   \caption{$Spitzer$/IRAC $8\,\mu$m emission shown by the grayscale image
    with $10\sigma_I$ $850$\,$\mu$m contours in red. The positions of (candidate) YSOs
    are overplotted in blue \citep[2MASS:][]{Cutri+2003} and green \citep[WISE:][]{Cambresy+2013}.
    YSOs detected in X-ray by Chandra are overplotted in yellow/red \citep{Wang+2009, Broos+2013}.  
   } 
   \label{fig:8mu}%
\end{figure}

Most of the clumpy and elongated $850$\,$\mu$m features seem to lie along emission at 
IRAC/MIPS wavelengths which look like infrared dark clouds (IRDCs).
For a combined image of $3.6-4.5-5.8\,\mu$m and at $24\,\mu$m see this approximate subregion 
in Figure~10 of \citet{Poulton+2008}, and our Fig~\ref{fig:8mu} for the $8\,\mu$m coverage.

In Figure~\ref{fig:8mu} the $10\sigma_I$ $850$\,$\mu$m contours correlate well with lower 
$8\,\mu$m surface brightness around the cluster positions E and REFL08. 
It appears that there are fewer YSOs in the SCUBA-2 contours with the darkest $8\,\mu$m
emission, however for the positions and physical properties of the earlier stages of 
dense star-forming cores and protostars, using {\it Herschel}/HOBYS data, see 
Bontemps et al.~(in prep.).

\section{Discussion} \label{sec:discuss}

\subsection{Dust masses} \label{sec:mass}

The total mass of a region is one indicator of its potential for star formation.
Submillimeter flux densities are routinely used to estimate molecular cloud 
masses using the following formula:

\begin{equation}
M = \frac{d^2 F_{\rm \nu}}{\kappa_{\rm \nu} B_{\rm \nu}(T_{\rm d})},
\label{eq:Mass}
\end{equation}

\noindent
where $d$ (1600\,pc) is the distance to the RMC, $F_{\rm \nu}$ is the total flux density at 850\,$\mu$m, 
$\kappa_{\rm \nu}$ is the dust mass opacity, and $B_{\rm \nu}(T_{\rm d})$ is the Planck function at 
dust temperature $T_{\rm d}$. 

We follow other BISTRO papers and the method of \citet{Beckwith+1990}
and formulate $\kappa_{\rm \nu}$ as 
$0.1(\nu/{\rm THz})^{\beta}$\,cm$^{2}$\,g$^{-1}$, assuming a standard dust-to-gas 
ratio of 1:100. The dust emissivity index, $\beta$, has been fixed to 2 
\citep[e.g.,][]{Hildebrand1983, Roy+2014, Pattle+2015}. 

\begin{figure}[!t]
 \begin{center}
 \begin{minipage}{1.\linewidth}
 \resizebox{1.0\hsize}{!}{\includegraphics[angle=0]{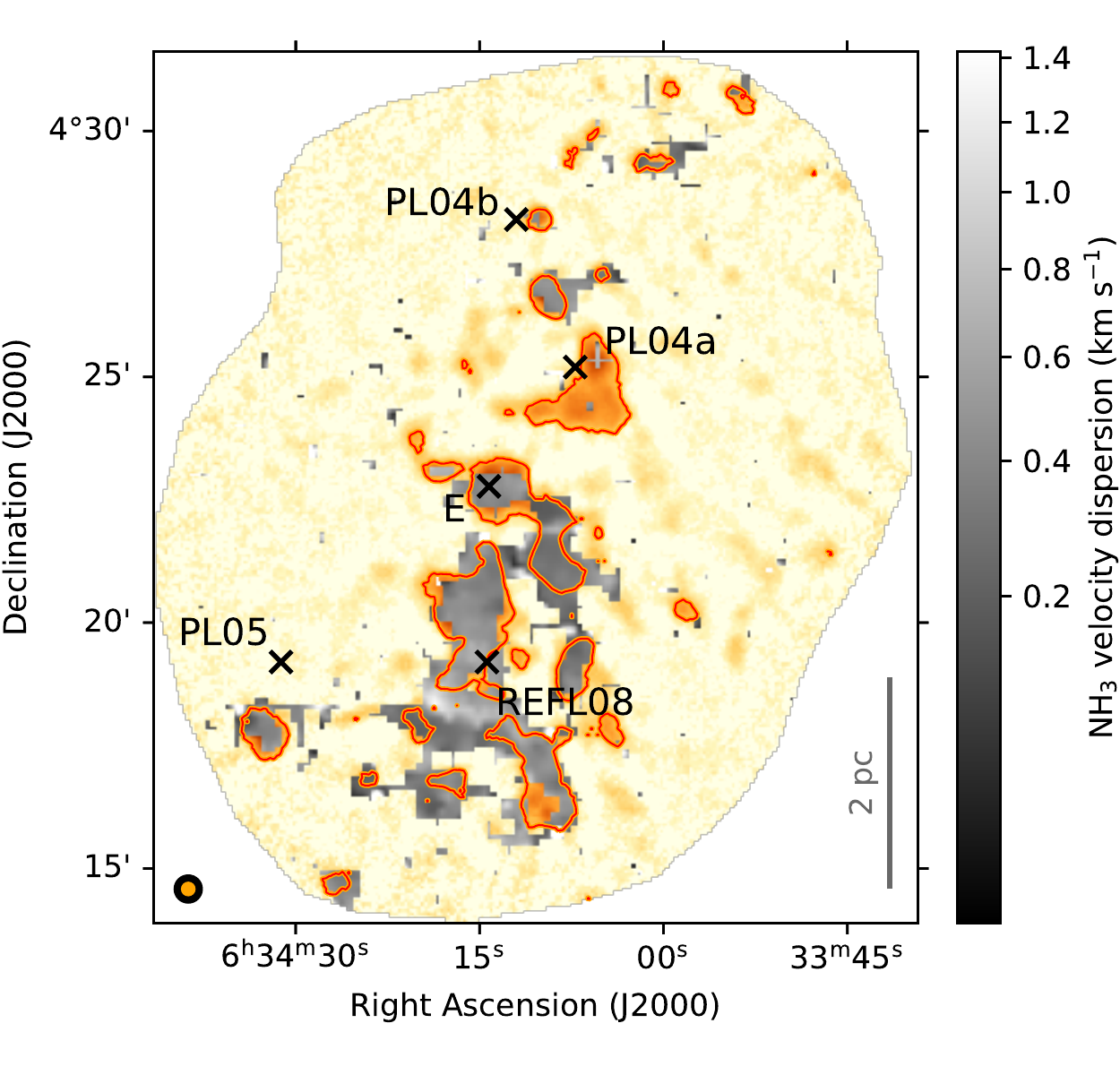}} 
\end{minipage}
 \end{center}
   \caption{Background image is our $850$\,$\mu$m Stokes $I$ map, with red contours 
    at $10 \sigma_{I} = 0.13$~mJy/arcsec$^2$. In grayscale the NH$_{\rm 3}$(1,1) velocity
    dispersions are overplotted from the KEYSTONE Survey \citep{Keown+2019}.
    In the lower left corner the spatial resolutions of the JCMT $850$\,$\mu$m ($\sim$14\arcsec)
    and the ammonia data (32\arcsec) are marked as orange and black circles, respectively. 
    }   
   \label{fig:NH3}%
\end{figure}

Within our mapped field ($\sim 12.9\arcmin \times 16.1\arcmin$, or $\sim 6 \times 7.5$\,pc), 
which is the whole region in Fig.~\ref{fig:NH3}, 
we derive a mass of $\sim 174\, M_\odot$ for the RMC center.
This mass includes $\sim 15\, M_\odot$ for the ring region, and $\sim 84\, M_\odot$
for the southern field, both estimated within their boxes outlined in Fig.~\ref{fig:Imap}.
Within the region in the Stokes $I$ image where $I/\delta I > 10$ 
(see the contours, for example, in Figs.~\ref{fig:Imap} and \ref{fig:NH3}), 
the mass corresponds to $\sim 41\, M_\odot$. 

For these masses we used a median $T_{\rm d}$ for each field that we estimated from the 
{\it Herschel} dust temperature image (see below, and also Ladjelate et al., in prep.). 
Assuming a typical factor of 2 uncertainty on the mass, we claim that there may be 
a few hundred solar masses of material in the densest regions probed by the JCMT. 

Looking at our mapped field in {\it Herschel}/HOBYS H$_{\rm 2}$ column density data 
(Ladjelate et al., in prep.), 
the total mass was derived as in, for example, \citet[][]{Konyves+2015, Konyves+2020} 
and resulted in $\sim 9.4 \times 10^{3}\, M_\odot$.
This {\it Herschel}-mass is about 2.5-times as much as 
that of the dense molecular gas 
material available in Orion~B \citep{Konyves+2020},
in which low- to high-mass star formation is also occurring. 
At the same time, it represents only about $7$\%
of the total mass of the whole Rosette Molecular Cloud 
region seen by {\it Herschel} 
(see this coverage in Fig.~\ref{fig:bigView}).

In order to make a comparison between a ground-based instrument, such as SCUBA-2,
and a satellite, such as {\it Herschel}, it is necessary to take account of the 
very extended surface brightness seen by {\it Herschel}, to which SCUBA-2 is insensitive. 
To make such a 
comparison between our SCUBA-2 masses and {\it Herschel}
masses we have taken the {\it Herschel} $250\,\mu$m data which have a similar resolution 
($18.2$\arcsec) to the SCUBA-2 data. In our effective mapped area we have selected a 
relatively empty $\sim 2$\arcmin-diameter region, where we measured the median 
surface brightness in the $250\,\mu$m map and used this offset to subtract the 
large-scale emission from the latter map.       
When we measure the remaining flux density at $250\,\mu$m within our field (see the whole 
region in Fig.~\ref{fig:NH3}) and use Eq.~\ref{eq:Mass}, we obtain a mass of $\sim 238\, M_\odot$. 
This shows good agreement (within 30\%) with the mass we derived from SCUBA-2, above. 

We note, however, that SCUBA-2's spatial filtering is more complicated than removing 
a zero-level offset; with different amounts of emission levels being removed at different 
scales up to 5\arcmin. In addition, the choice of the dust emissivity index, $\beta$, or
the dust opacity, $\kappa_{\rm \nu}$, may also adjust the result of this comparison.   

The brightest 850-$\mu$m emission pixels can be found in the clump south of 
PL05 (see e.g., Fig.~\ref{fig:Imap}),
where the corresponding average dust temperature and column densities give $18\,$K, 
and $N_{\rm H_2} \sim 2.3 \times 10^{22}$\,cm$^{-2}$, respectively.
Apart from this, the one other `hot-spot' in our image is the gemstone head of  
the ring with $T \sim 19$\,K. 
These two warmer spots correlate with locations of stronger $8\,\mu$m emission 
(see Fig.~\ref{fig:8mu}), and have somewhat lower column density than the colder 
($15-16\,$K) southern filamentary clumps. 
Indeed, colder areas tend to have higher column densities,  
where the absence of thermal heating and pressure support allow the matter to
become more compact and eventually collapse into stars.
For a comparison of the distribution of our ``a priori'' cold and dense $850$\,$\mu$m emission 
and the hot $8$\,$\mu$m PAH emission, see Fig.~\ref{fig:8mu}.  

\subsection{Magnetic Field Strength and Stability} \label{sec:strength}

The most commonly used method to infer the field strength from polarized dust emission is the 
Davis-Chandrasekhar-Fermi (DCF) technique \citep{Davis1951, Chandrasekhar+Fermi1953} 
-- see also work by \citet{Houde+2016} and \citet{Pattle+2017},
and the discussion of its caveats and limitations in 
\citet[][and references therein]{PattleFissel2019}.
This method estimates the B-field strength by comparing the dispersion in the 
polarization orientation (assumed to be a measure of the non-uniform B-field)
with the dispersion in LOS velocity (assumed to be a measure of the non-thermal
motions of the gas).
This method assumes small-scale non-thermal motions, thus should not be applied 
under super-Alfv\'enic turbulent conditions, i.e., when $\mathcal{M}_{\rm A} \gg 1$, 
where $\mathcal{M}_{\rm A}$ is the Alfv\'en Mach number. 
Following \citet{Pattle+2020b}, it can be expressed as  
\begin{equation}
  \mathcal{M}_{\rm A} \approx 3.5 \times 10^{-2} \sigma_{\theta},
\label{eq:MA}
\end{equation}
where $\sigma_{\theta}$ is the polarization angle dispersion in degrees.

When this condition holds, the POS magnetic field strength in ${\rm \mu G}$ can thus 
be estimated using the equation
\begin{equation}
\begin{split}
  B_{\rm POS} & \approx Q' \sqrt{4 \pi \rho}~\frac{\sigma_v}{\sigma_{\theta}} 
                \approx 9.3 \sqrt{n_{\rm H_2}}~\frac{\Delta v_{\rm NT}}{\sigma_{\theta}} ,
\end{split}
\label{eq:Bpos}
\end{equation}
where $\rho$ is the mean density of the cloud or subregion in g\,cm$^{-3}$;
$\sigma_v$ is the velocity dispersion in km\,s$^{-1}$;
$n_{\rm H_2}$ is the hydrogen molecule number density in cm$^{-3}$; and 
$\Delta v_{\rm NT}$ is the non-thermal line width in km\,s$^{-1}$.
In order to simplify the left-hand side of Eq.~\ref{eq:Bpos} and arrive at the 
right-hand side formulation, we followed \citet{Crutcher+2004}.
Under strong B-field conditions ($\sigma_{\theta} \lesssim 25\degr$) the factor of
$Q' = 0.5$ can provide a somewhat more accurate estimate of the POS field strength 
\citep{Ostriker+2001, Lai+2002}
that \citet{Crutcher+2004} also find to be a reasonable value in dense, 
self-gravitating cores and filaments with expected little field substructure.
We again refer the reader to \citet[][and references therein]{PattleFissel2019}
for the discussion on the telescope beam effects that are parameterized in this 
correction factor, $Q'$.   

We used our Rosette POL-2 data to estimate the polarization angle dispersion. 
Corresponding molecular line observations of NH$_{\rm 3}$(1,1) from the KEYSTONE Survey 
\citep[][Di Francesco et al., in prep.]{Keown+2019} were used for deriving the 
line widths, and we calculated the H$_{\rm 2}$ volume density from 
{\it Herschel}-derived masses.

First, we discuss the dispersion of polarization angles, as this parameter is estimated 
directly from our POL-2 data. 
In order to select the magnetic half-vectors, and so the subregion toward which we can derive 
the field strength, we considered that we need a statistically significant number of half-vectors 
at $14\arcsec$~binning, and we need there to be
molecular line observations in the same location. 
See the coverage of the NH$_{\rm 3}$(1,1) data we used in the RMC in Fig.~\ref{fig:NH3}, 
where the ammonia velocity dispersion is over-plotted on the $850$\,$\mu$m Stokes $I$ map.
  
\begin{figure}[!hhh]
 \begin{center}
 \begin{minipage}{1.0\linewidth}
 \hspace{-1mm}
 \resizebox{1.01\hsize}{!}{\includegraphics[angle=0]{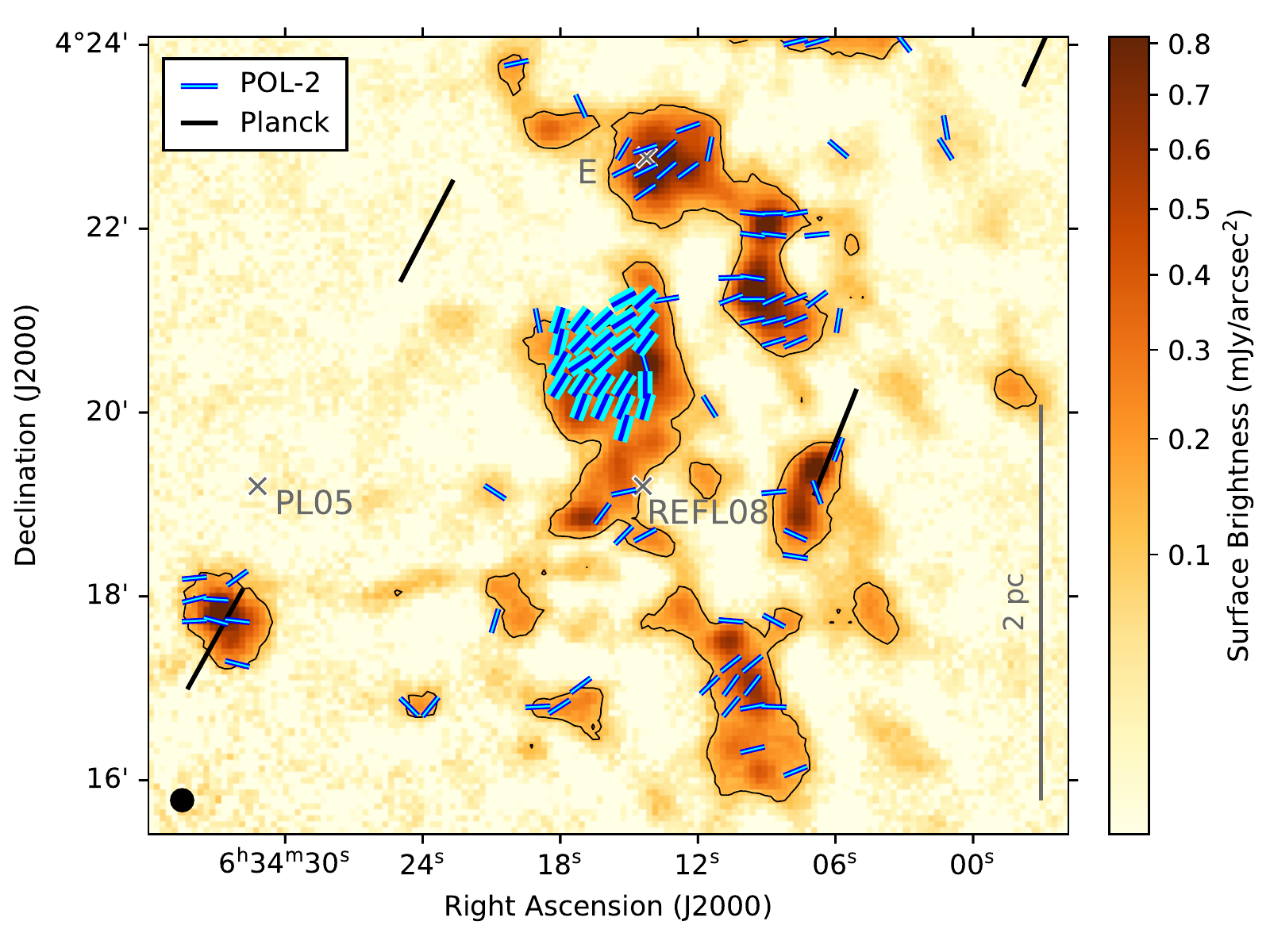}} 
\end{minipage}
 \end{center}
   \caption{Same as the right-hand 
   panel of Fig.~\ref{fig:pol14as} with highlighted B-field half-vectors
   (in thick sections) in which subregion it was 
   possible to derive the magnetic field strength (see text for details).          
   } 
   \label{fig:Bpos}%
\end{figure}

Taking the above considerations into account led us to only one subregion for 
which B-field strength estimates were possible. 
This subregion is indicated by the thick B-field half-vectors in Fig.~\ref{fig:Bpos} 
to the north of REFL08.

We calculated the standard deviation of the polarization angles in the selected 
group of 25 half-vector segments.
This simple measure for the polarization dispersion is only possible because 
a priori we chose segments which spread over a relatively narrow range in orientations. 
The uncertainty of the angle dispersion originates from the median angle 
uncertainty of the selected individual vectors (see Table~\ref{tab:Bpos}). 

The most suitable molecular line data from which we could estimate the line widths 
were obtained from the KEYSTONE Survey 
\citep[][Di Francesco et al., in prep.]{Keown+2019}, a large project  
on the 100-m Green Bank Telescope. This survey is mapping ammonia emission in several giant 
molecular clouds in order to characterise massive star formation. 
The spatial resolution of the data cubes is 32\arcsec~projected on 8.8\arcsec pixel scale,
with a spectral resolution of 0.07~km~s$^{-1}$ \citep{Keown+2019}.

Ammonia molecules are less prone to freezing out than CO at high densities, and their
emission lines normally stay 
optically thin. They can probe deep layers of molecular clouds, and are typically associated with densities 
above $\sim 10^{4}$\,cm$^{-3}$ \citep[e.g.,][]{Benson+Myers1989}.

The FWHM ammonia line width, $\Delta v$, was calculated from the 
velocity dispersion (see Fig.~\ref{fig:NH3}) as $\Delta v = \sigma_v\sqrt{8~{\rm ln}2}$.
We then separated the non-thermal component $\Delta v_{\rm NT}$ in km\,s$^{-1}$ 
using a similar relation to eqn.~B8 of \citet{Pattle+2020b}. 

We estimated an average $n_{\rm H_2}$ volume number density from {\it Herschel} masses 
in the subregion which is defined by the selected half-vector segments highlighted in 
Fig.~\ref{fig:Bpos}.
The {\it Herschel} mass of this subfield was derived from $N_{\rm H_2}$ column densities 
as in Sect.~\ref{sec:mass}.
Then we calculated the volume density following, for example, \citet[][]{Pattle+2020b}:
\begin{equation}
  n_{\rm H_2} = \frac{M}{\mu m_{\rm H}} \frac{3}{4 \pi R^3},
\label{eq:nH2}
\end{equation}
%
where $\mu = 2.8$ is the mean molecular weight per ${\rm H_2}$ molecule, $m_{\rm H}$ is 
the hydrogen atom mass, and $R$ is the radius of a circle with equivalent area 
of the subfield occupied by the selected vectors.

The derived parameters and their uncertainties relevant to the DCF analysis, along with the 
estimated Alfv\'en Mach number and field strength, are summarized in Table~\ref{tab:Bpos} 
for the subregion highlighted by the selected vectors in Fig.~\ref{fig:Bpos}.
%
\begin{table}[!hb]\small\setlength{\tabcolsep}{4.pt}
\centering
 \caption{ Measured and derived properties relevant to the DCF analysis in a subregion of the RMC (see Fig.~\ref{fig:Bpos}). 
          Abbreviations: {\it tot} means total, {\it ave} means average, and {\it equiv} means equivalent value 
          over the subregion.
          See text for details. }
 \label{tab:Bpos}      
  {\renewcommand{\arraystretch}{1.2}        
   \begin{tabular}{|l|c|}  
   \hline
    \hline
    Property						& Value \\
    \hline\hline
    Pol. angle dispersion, $\sigma_{\theta}$			& 15.43\degr $\pm$ 5.33\degr \\

    Ammonia line width, $\Delta v_{\rm NT}$			& 0.97 $\pm$ 0.14 km s$^{-1}$ \\

    H$_{\rm 2}$ column density, $N_{\rm H_2}^{\rm tot}$		& (7.65 $\pm$ 1.53) $\times$ 10$^{24}$ cm$^{-2}$ \\

    H$_{\rm 2}$ column density, $N_{\rm H_2}^{\rm ave}$		& (2.45 $\pm$ 0.49) $\times$ 10$^{22}$ cm$^{-2}$ \\

    Mass of subregion, $M^{\rm tot}$				& 165 $\pm$ 33 $M_\odot$ \\

    Radius of subregion, $R^{\rm equiv}$			& 0.31 pc \\

    H$_{\rm 2}$ number density, $n_{\rm H_2}^{\rm ave}$ 	& (1.90 $\pm$ 0.38) $\times$ 10$^4$ cm$^{-3}$ \\
    \hline

    Alfv\'en Mach number, $\mathcal{M}_{\rm A}$			& 0.54 $\pm$ 0.19 \\

    B-field strength, $B_{\rm POS}$ 				& 80 $\pm$ 30 $\mu$G \\

    Mass-to-flux ratio, $\lambda$ 				& 2.3 $\pm$ 1.0 \\
   \hline\hline
   \end{tabular}
  }
\end{table}

The uncertainties on the column densities, and so on the mass and volume densities, 
were taken to be 20\% in these calculation, in order to avoid the propagation of the typical 
factor of about 2 systematic errors mainly due to the uncertainties in the dust opacity law. 
For a more subtle treatment of these systematic errors in the DCF analysis see \citet[][]{Pattle+2020b}.
With other derived properties we use the quadratic addition of errors. 
Our magnetic field strengths with the DCF method, with assumed resolved 
structure at the clump level, are found to be typically correct to within a factor of 2, 
based on numerical simulations by \citet{Heitsch+2001}. In this factor, only uncertainties 
originating from the polarization observations (i.e., resolution effects) are considered.

We find the B-field strength toward a dense clump and its outskirts to be 
$80 \pm 30~\mu$G which corresponds to the range of $50 - 110$\,$\mu$G.
These values are local and thus not clearly representative of the larger environment.
The difference in the observed spatial scales and the fact that SCUBA-2 can 
resolve higher densities are the reasons why we cannot meaningfully compare our results with 
the {\it Planck}-found $B_{\rm LOS} \sim 3\,\mu$G in the Rosette Nebula \citep{PlanckCollabXXXIV+2016}.

Given that the diffuse ISM shows a well-defined median magnetic field strength 
of $6.0 \pm 1.8$\,$\mu$G \citep{HeilesTroland2005}, our field is  
at least an order of magnitude stronger. 
The B-field strength in this clump and its surroundings seems to be comparable 
to that of the nearby starless core L1689B \citep{Pattle+2020b}, and also 
to those in the northern and southern parts of the G34 IRDC at a distance of 
$\sim 3.7$\,kpc \citep{Soam+2019}.
The configuration of our selected subregion for the $B_{\rm POS}$ calculations
looks more similar to the geometry of the northern part of the elongated
G34 IRDC in that they both contain dense core(s), which are probably 
already protostellar, as well their surrounding environment.  
Our magnetic field values are typically comparable to, or weaker than those found 
in other IRDCs \citep[e.g.,][]{Pillai+2015, Liu+2018}.

Magnetism is an important component of the ISM, however it is the ratio of mass 
and magnetic flux that can determine the relative importance of magnetic and 
gravitational forces, so the stability of the investigated region. 
We estimated the mass-to-flux ratio, $\lambda$, with the formula given by 
\citet{Crutcher+2004}:
\begin{equation}
  \lambda = 7.6 \times 10^{-21} \frac{N_{\rm H_2}}{B_{\rm POS}},
\label{eq:lam}
\end{equation}
where the average H$_{\rm 2}$ column density is assumed in cm$^{-2}$, and 
$B_{\rm POS}$ in $\mu$G.  

With the values in Table~\ref{tab:Bpos} we derive $\lambda = 2.3 \pm 1.0$,
which is higher than the critical value $\lambda = 1$, and suggests that
the investigated subregion is gravitationally unstable; the magnetically 
supercritical B-field is not strong enough to prevent gravitational collapse.
This result is not surprising, given that we are in the actively star-forming 
central ridge of the RMC that is producing high-mass stars.

\section{Conclusions} \label{sec:conc}

As part of the BISTRO-2 survey using SCUBA-2/POL-2 at the JCMT, we have presented $850$\,$\mu$m 
polarization observations toward the center of the Rosette Molecular Cloud 
within an effective area of $\sim 6 \times 7.5$\,pc
at $\sim 1.6$\,kpc distance. 
Our main results and conclusions are summarized as follows:
%
\begin{enumerate}
  \item In our analysis, we used polarization vector selection criteria of
  Stokes $I > 0$, $I/\delta I > 10$, 
  $p/\delta p > 3$, and $\delta p < 5\%$,
  which gave us $152$ vectors at $14$\arcsec~sampling. 
  \item We assessed the dust grain alignment through the dependence of polarization fraction 
  on total intensity which shows a $p \propto I^{\rm -\alpha}$ relation. We find 
  $\alpha = 0.49 \pm 0.08$ which suggests that a significant
  fraction of the dust grains remain aligned 
  with respect to the magnetic field in the highest observed densities. 
  \item In the north of our region the $850\,\mu$m image reveals a ring-like structure with a diameter 
  of $\sim 1$\,pc. Its emission is strongest in the south-west. We refer to this overall structure as 
  a ``gemstone ring'', which is seen to be filled with 
  {\it Spitzer} emission at $3.6-24$\,$\mu$m. This short-to-long wavelength emission 
  forms a gradient  
  that, in places, appears to sit on the SCUBA-2 clumps making up the rim of the bubble wall. 
  The B-field seems to partially trace the circumference of the ring, which turns almost 
  perpendicular to it in the western part where there is a brighter clump. 
  \item In the southern part of the RMC center, the SCUBA-2 data shows clumpy emission with connecting 
  filaments that follow IRDCs.
  Here, the POL-2 B-field geometry appears bimodal with some clumps well aligned with the large-scale 
  {\it Planck} field and some clumps nearly perpendicular.   
  \item From the $850$\,$\mu$m flux densities within our effective mapped area we derive a mass of 
  $\sim 174\, M_\odot$. 
  We compare our results with large-scale emission-subtracted {\it Herschel} $250$\,$\mu$m-masses 
  and find that the two values agree to within $30$\%.   
  \item Using the DCF technique we estimate the POS B-field strength in one subregion of 
  our field, toward a dense clump and its outskirts. 
  We find a value of $80 \pm 30$\,$\mu$G that is typically comparable or weaker than the field 
  strength in IRDCs. 
  \item The mass-to-flux ratio ($\lambda = 2.3 \pm 1.0$) of this subfield suggests that
  the B-field is not sufficiently strong to prevent gravitational collapse. 
\end{enumerate}

\acknowledgments
We thank the anonymous reviewer for their helpful suggestions.
V.K.~and D.W.-T.~acknowledge Science and Technology Facilities Council (STFC) support under grant 
number ST/R000786/1.
P.N.D.~is funded by the Vietnam National Foundation for Science and Technology Development (NAFOSTED) 
under grant No.~103.99-2019.368.
C.L.H.H.~acknowledges the support of the NAOJ Fellowship and JSPS KAKENHI grants 18K13586 and 20K14527.
W.K.~was supported by the New Faculty Startup Fund from Seoul National University.
C.W.L. is supported by the Basic Science Research Program through the NRF funded
by the Ministry of Education, Science and Technology (NRF-2019R1A2C1010851).
K.Q.~is partially supported by National Key R\&D Program of China No.~2017YFA0402600, and acknowledges 
the National Natural Science Foundation of China (NSFC) grant U1731237.
The James Clerk Maxwell Telescope is operated by the East Asian Observatory on behalf of the 
National Astronomical Observatory of Japan; Academia Sinica Institute of Astronomy and Astrophysics;  
the Korea Astronomy and Space Science Institute (KASI); the Operation, Maintenance and Upgrading Fund 
for Astronomical Telescopes and Facility Instruments, budgeted from the Ministry of Finance (MOF) of
China and administrated by the Chinese Academy of Sciences (CAS); and the National Key R\&D Program 
of China (No. 2017YFA0402700). Additional funding support is provided by the Science and Technology  
Facilities Council of the United Kingdom and participating universities in the United Kingdom and 
Canada. SCUBA-2 and POL-2 were built through grants from the Canada Foundation for Innovation.  
This research used the facilities of the Canadian Astronomy Data Centre operated by the National 
Research Council of Canada with the support of the Canadian Space Agency.   
This research has made use of the SIMBAD database, and the ``Aladin sky atlas'' developed and operated
at CDS, Strasbourg Observatory, France.
This research has also made use of NASA's Astrophysics Data System Bibliographic Services.


%
\facilities{
JCMT(SCUBA-2, POL-2), Herschel(SPIRE, PACS), SPITZER(IRAC, MIPS)
            }


\software{
APLpy \citep{Robitaille+Bressert2012}, 
Astropy (Astropy Collaboration et al. 2013),  
Numpy \citep{Harris+2020},
Matplotlib \citep{Hunter2007}, 
Starlink \citep{Berry+2005, Chapin+2013}
          }


\bibliography{./BISTRO2_Rosette_konyves_2021_ApJsubm_resubm_fin}{}

\begin{thebibliography}{}
\expandafter\ifx\csname natexlab\endcsname\relax\def\natexlab#1{#1}\fi

\bibitem[{{Anderson} {et~al.}(2011){Anderson}, {Bania}, {Balser}, \&
  {Rood}}]{Anderson+2011}
{Anderson}, L.~D., {Bania}, T.~M., {Balser}, D.~S., \& {Rood}, R.~T. 2011,
  \apjs, 194, 32

\bibitem[{{Arthur} {et~al.}(2011){Arthur}, {Henney}, {Mellema}, {de Colle}, \&
  {V{\'a}zquez-Semadeni}}]{Arthur+2011}
{Arthur}, S.~J., {Henney}, W.~J., {Mellema}, G., {de Colle}, F., \&
  {V{\'a}zquez-Semadeni}, E. 2011, \mnras, 414, 1747

\bibitem[{{Beckwith} {et~al.}(1990){Beckwith}, {Sargent}, {Chini}, \&
  {Guesten}}]{Beckwith+1990}
{Beckwith}, S.~V.~W., {Sargent}, A.~I., {Chini}, R.~S., \& {Guesten}, R. 1990,
  \aj, 99, 924

\bibitem[{{Bell} {et~al.}(2013){Bell}, {Naylor}, {Mayne}, {Jeffries}, \&
  {Littlefair}}]{Bell+2013}
{Bell}, C. P.~M., {Naylor}, T., {Mayne}, N.~J., {Jeffries}, R.~D., \&
  {Littlefair}, S.~P. 2013, \mnras, 434, 806

\bibitem[{{Benson} \& {Myers}(1989)}]{Benson+Myers1989}
{Benson}, P.~J., \& {Myers}, P.~C. 1989, \apjs, 71, 89

\bibitem[{{Berry} {et~al.}(2005){Berry}, {Gledhill}, {Greaves}, \&
  {Jenness}}]{Berry+2005}
{Berry}, D.~S., {Gledhill}, T.~M., {Greaves}, J.~S., \& {Jenness}, T. 2005, in
  Astronomical Society of the Pacific Conference Series, Vol. 343, Astronomical
  Polarimetry: Current Status and Future Directions, ed. A.~{Adamson},
  C.~{Aspin}, C.~{Davis}, \& T.~{Fujiyoshi}, 71

\bibitem[{{Bica} {et~al.}(2003){Bica}, {Dutra}, \& {Barbuy}}]{Bica+2003}
{Bica}, E., {Dutra}, C.~M., \& {Barbuy}, B. 2003, \aap, 397, 177

\bibitem[{{Broos} {et~al.}(2013){Broos}, {Getman}, {Povich}, {Feigelson},
  {Townsley}, {Naylor}, {Kuhn}, {King}, \& {Busk}}]{Broos+2013}
{Broos}, P.~S., {Getman}, K.~V., {Povich}, M.~S., {et~al.} 2013, \apjs, 209, 32

\bibitem[{{Cambr{\'e}sy} {et~al.}(2013){Cambr{\'e}sy}, {Marton}, {Feher},
  {T{\'o}th}, \& {Schneider}}]{Cambresy+2013}
{Cambr{\'e}sy}, L., {Marton}, G., {Feher}, O., {T{\'o}th}, L.~V., \&
  {Schneider}, N. 2013, \aap, 557, A29

\bibitem[{{Chandrasekhar} \& {Fermi}(1953)}]{Chandrasekhar+Fermi1953}
{Chandrasekhar}, S., \& {Fermi}, E. 1953, \apj, 118, 113

\bibitem[{{Chapin} {et~al.}(2013){Chapin}, {Berry}, {Gibb}, {Jenness}, {Scott},
  {Tilanus}, {Economou}, \& {Holland}}]{Chapin+2013}
{Chapin}, E.~L., {Berry}, D.~S., {Gibb}, A.~G., {et~al.} 2013, \mnras, 430,
  2545

\bibitem[{{Chen} {et~al.}(2017){Chen}, {Jiang}, {Tamura}, {Kwon}, \&
  {Roman-Lopes}}]{Chen+2017}
{Chen}, Z., {Jiang}, Z., {Tamura}, M., {Kwon}, J., \& {Roman-Lopes}, A. 2017,
  \apj, 838, 80

\bibitem[{{Costa} {et~al.}(2016){Costa}, {Spangler}, {Sink}, {Brown}, \&
  {Mao}}]{Costa+2016}
{Costa}, A.~H., {Spangler}, S.~R., {Sink}, J.~R., {Brown}, S., \& {Mao}, S.~A.
  2016, \apj, 821, 92

\bibitem[{{Coud{\'e}} {et~al.}(2019){Coud{\'e}}, {Bastien}, {Houde}, {Sadavoy},
  {Friesen}, {Di Francesco}, {Johnstone}, {Mairs}, {Hasegawa}, {Kwon}, {Lai},
  {Qiu}, {Ward-Thompson}, {Berry}, {Chen}, {Fiege}, {Franzmann}, {Hatchell},
  {Lacaille}, {Matthews}, {Moriarty-Schieven}, {Pon}, {Andr{\'e}},
  {Arzoumanian}, {Aso}, {Byun}, {Eswaraiah}, {Chen}, {Chen}, {Ching}, {Cho},
  {Choi}, {Chrysostomou}, {Chung}, {Doi}, {Drabek-Maunder}, {Dowell}, {Eyres},
  {Falle}, {Friberg}, {Fuller}, {Furuya}, {Gledhill}, {Graves}, {Greaves},
  {Griffin}, {Gu}, {Hayashi}, {Hoang}, {Holland }, {Inoue}, {Inutsuka},
  {Iwasaki}, {Jeong}, {Kanamori}, {Kataoka}, {Kang}, {Kang}, {Kang},
  {Kawabata}, {Kemper}, {Kim}, {Kim}, {Kim}, {Kim}, {Kim}, {Kim}, {Kirk},
  {Kobayashi}, {Koch}, {Kwon}, {Lee}, {Lee}, {Lee}, {Li}, {Li}, {Li}, {Liu},
  {Liu}, {Liu}, {Liu}, {van Loo}, {Lyo}, {Matsumura}, {Nagata}, {Nakamura},
  {Nakanishi}, {Ohashi}, {Onaka}, {Parsons}, {Pattle}, {Peretto}, {Pyo},
  {Qian}, {Rao}, {Rawlings}, {Retter}, {Richer}, {Rigby}, {Robitaille},
  {Saito}, {Savini}, {Scaife}, {Seta}, {Shinnaga}, {Soam}, {Tamura}, {Tang},
  {Tomisaka}, {Tsukamoto}, {Wang}, {Wang}, {Whitworth}, {Yen}, {Yoo}, {Yuan},
  {Zenko}, {Zhang}, {Zhang}, {Zhou}, {Zhu}, \& {B-fields In STar-forming
  Regions Observations (BISTRO Collaboration}}]{Coude+2019}
{Coud{\'e}}, S., {Bastien}, P., {Houde}, M., {et~al.} 2019, \apj, 877, 88

\bibitem[{{Crutcher} {et~al.}(2004){Crutcher}, {Nutter}, {Ward-Thompson}, \&
  {Kirk}}]{Crutcher+2004}
{Crutcher}, R.~M., {Nutter}, D.~J., {Ward-Thompson}, D., \& {Kirk}, J.~M. 2004,
  \apj, 600, 279

\bibitem[{{Currie} {et~al.}(2014){Currie}, {Berry}, {Jenness}, {Gibb}, {Bell},
  \& {Draper}}]{Currie+2014}
{Currie}, M.~J., {Berry}, D.~S., {Jenness}, T., {et~al.} 2014, in Astronomical
  Society of the Pacific Conference Series, Vol. 485, Astronomical Data
  Analysis Software and Systems XXIII, ed. N.~{Manset} \& P.~{Forshay}, 391

\bibitem[{{Cutri} \& {et al.}(2012)}]{Cutri+2012}
{Cutri}, R.~M., \& {et al.} 2012, VizieR Online Data Catalog, II/311

\bibitem[{{Cutri} {et~al.}(2003){Cutri}, {Skrutskie}, {van Dyk}, {Beichman},
  {Carpenter}, {Chester}, {Cambresy}, {Evans}, {Fowler}, {Gizis}, {Howard},
  {Huchra}, {Jarrett}, {Kopan}, {Kirkpatrick}, {Light}, {Marsh}, {McCallon},
  {Schneider}, {Stiening}, {Sykes}, {Weinberg}, {Wheaton}, {Wheelock}, \&
  {Zacarias}}]{Cutri+2003}
{Cutri}, R.~M., {Skrutskie}, M.~F., {van Dyk}, S., {et~al.} 2003, VizieR Online
  Data Catalog, II/246

\bibitem[{{Davis}(1951)}]{Davis1951}
{Davis}, L. 1951, Physical Review, 81, 890

\bibitem[{{Dent} {et~al.}(2009){Dent}, {Hovey}, {Dewdney}, {Burgess}, {Willis},
  {Lightfoot}, {Jenness}, {Leech}, {Matthews}, {Heyer}, \&
  {Poulton}}]{Dent+2009}
{Dent}, W.~R.~F., {Hovey}, G.~J., {Dewdney}, P.~E., {et~al.} 2009, \mnras, 395,
  1805

\bibitem[{{Di Francesco} {et~al.}(2008){Di Francesco}, {Johnstone}, {Kirk},
  {MacKenzie}, \& {Ledwosinska}}]{DiFrancesco+2008}
{Di Francesco}, J., {Johnstone}, D., {Kirk}, H., {MacKenzie}, T., \&
  {Ledwosinska}, E. 2008, \apjs, 175, 277

\bibitem[{{di Francesco} {et~al.}(2010){di Francesco}, {Sadavoy}, {Motte},
  {Schneider}, {Hennemann}, {Csengeri}, {Bontemps}, {Balog}, {Zavagno},
  {Andr{\'e}}, {Saraceno}, {Griffin}, {Men'shchikov}, {Abergel}, {Baluteau},
  {Bernard}, {Cox}, {Deharveng}, {Didelon}, {di Giorgio}, {Hargrave}, {Huang},
  {Kirk}, {Leeks}, {Li}, {Marston}, {Martin}, {Minier}, {Molinari}, {Olofsson},
  {Persi}, {Pezzuto}, {Russeil}, {Sauvage}, {Sibthorpe}, {Spinoglio}, {Testi},
  {Teyssier}, {Vavrek}, {Ward-Thompson}, {White}, {Wilson}, \&
  {Woodcraft}}]{DiFrancesco+2010}
{di Francesco}, J., {Sadavoy}, S., {Motte}, F., {et~al.} 2010, \aap, 518, L91

\bibitem[{{Friberg} {et~al.}(2016){Friberg}, {Bastien}, {Berry}, {Savini},
  {Graves}, \& {Pattle}}]{Friberg+2016}
{Friberg}, P., {Bastien}, P., {Berry}, D., {et~al.} 2016, in Society of
  Photo-Optical Instrumentation Engineers (SPIE) Conference Series, Vol. 9914,
  Millimeter, Submillimeter, and Far-Infrared Detectors and Instrumentation for
  Astronomy VIII, 991403

\bibitem[{{Friberg} {et~al.}(2018){Friberg}, {Berry}, {Savini}, {Bintley},
  {Dempsey}, {Graves}, \& {Parsons}}]{Friberg+2018}
{Friberg}, P., {Berry}, D., {Savini}, G., {et~al.} 2018, in Society of
  Photo-Optical Instrumentation Engineers (SPIE) Conference Series, Vol. 10708,
  Millimeter, Submillimeter, and Far-Infrared Detectors and Instrumentation for
  Astronomy IX, ed. J.~{Zmuidzinas} \& J.-R. {Gao}, 107083M

\bibitem[{{Guillet} {et~al.}(2020){Guillet}, {Girart}, {Maury}, \&
  {Alves}}]{Guillet+2020}
{Guillet}, V., {Girart}, J.~M., {Maury}, A.~J., \& {Alves}, F.~O. 2020, \aap,
  634, L15

\bibitem[{Harris {et~al.}(2020)Harris, Millman, van~der Walt, Gommers,
  Virtanen, Cournapeau, Wieser, Taylor, Berg, Smith, Kern, Picus, Hoyer, van
  Kerkwijk, Brett, Haldane, del R{'{\i}}o, Wiebe, Peterson,
  G{'{e}}rard-Marchant, Sheppard, Reddy, Weckesser, Abbasi, Gohlke, \&
  Oliphant}]{Harris+2020}
Harris, C.~R., Millman, K.~J., van~der Walt, S.~J., {et~al.} 2020, Nature, 585,
  357

\bibitem[{{Heiles} \& {Troland}(2005)}]{HeilesTroland2005}
{Heiles}, C., \& {Troland}, T.~H. 2005, \apj, 624, 773

\bibitem[{{Heitsch} {et~al.}(2001){Heitsch}, {Zweibel}, {Mac Low}, {Li}, \&
  {Norman}}]{Heitsch+2001}
{Heitsch}, F., {Zweibel}, E.~G., {Mac Low}, M.-M., {Li}, P., \& {Norman}, M.~L.
  2001, \apj, 561, 800

\bibitem[{{Hennemann} {et~al.}(2010){Hennemann}, {Motte}, {Bontemps},
  {Schneider}, {Csengeri}, {Balog}, {di Francesco}, {Zavagno}, {Andr{\'e}},
  {Men'shchikov}, {Abergel}, {Ali}, {Baluteau}, {Bernard}, {Cox}, {Didelon},
  {di Giorgio}, {Griffin}, {Hargrave}, {Hill}, {Horeau}, {Huang}, {Kirk},
  {Leeks}, {Li}, {Marston}, {Martin}, {Molinari}, {Nguyen Luong}, {Olofsson},
  {Persi}, {Pezzuto}, {Russeil}, {Saraceno}, {Sauvage}, {Sibthorpe},
  {Spinoglio}, {Testi}, {Ward-Thompson}, {White}, {Wilson}, \&
  {Woodcraft}}]{Hennemann+2010}
{Hennemann}, M., {Motte}, F., {Bontemps}, S., {et~al.} 2010, \aap, 518, L84

\bibitem[{{Hensberge} {et~al.}(2000){Hensberge}, {Pavlovski}, \&
  {Verschueren}}]{Hensberge+2000}
{Hensberge}, H., {Pavlovski}, K., \& {Verschueren}, W. 2000, \aap, 358, 553

\bibitem[{{Hildebrand}(1983)}]{Hildebrand1983}
{Hildebrand}, R.~H. 1983, QJRAS, 24, 267

\bibitem[{{Hoang} \& {Lazarian}(2016)}]{Hoang&Lazarian2016}
{Hoang}, T., \& {Lazarian}, A. 2016, \apj, 831, 159

\bibitem[{{Holland} {et~al.}(2013){Holland}, {Bintley}, {Chapin},
  {Chrysostomou}, {Davis}, {Dempsey}, {Duncan}, {Fich}, {Friberg}, {Halpern},
  {Irwin}, {Jenness}, {Kelly}, {MacIntosh}, {Robson}, {Scott}, {Ade},
  {Atad-Ettedgui}, {Berry}, {Craig}, {Gao}, {Gibb}, {Hilton}, {Hollister},
  {Kycia}, {Lunney}, {McGregor}, {Montgomery}, {Parkes}, {Tilanus}, {Ullom},
  {Walther}, {Walton}, {Woodcraft}, {Amiri}, {Atkinson}, {Burger}, {Chuter},
  {Coulson}, {Doriese}, {Dunare}, {Economou}, {Niemack}, {Parsons},
  {Reintsema}, {Sibthorpe}, {Smail}, {Sudiwala}, \& {Thomas}}]{Holland+2013}
{Holland}, W.~S., {Bintley}, D., {Chapin}, E.~L., {et~al.} 2013, \mnras, 430,
  2513

\bibitem[{{Houde} {et~al.}(2016){Houde}, {Hull}, {Plambeck}, {Vaillancourt}, \&
  {Hildebrand}}]{Houde+2016}
{Houde}, M., {Hull}, C. L.~H., {Plambeck}, R.~L., {Vaillancourt}, J.~E., \&
  {Hildebrand}, R.~H. 2016, \apj, 820, 38

\bibitem[{{Hull} \& {Zhang}(2019)}]{Hull+Zhang2019}
{Hull}, C. L.~H., \& {Zhang}, Q. 2019, Frontiers in Astronomy and Space
  Sciences, 6, 3

\bibitem[{{Hunter}(2007)}]{Hunter2007}
{Hunter}, J.~D. 2007, Computing in Science and Engineering, 9, 90

\bibitem[{{Jenness} {et~al.}(2013){Jenness}, {Chapin}, {Berry}, {Gibb},
  {Tilanus}, {Balfour}, {Tilanus}, \& {Currie}}]{Jenness+2013}
{Jenness}, T., {Chapin}, E.~L., {Berry}, D.~S., {et~al.} 2013, {SMURF:
  SubMillimeter User Reduction Facility}, , , ascl:1310.007

\bibitem[{{Jones} {et~al.}(2015){Jones}, {Bagley}, {Krejny}, {Andersson}, \&
  {Bastien}}]{Jones+2015}
{Jones}, T.~J., {Bagley}, M., {Krejny}, M., {Andersson}, B.~G., \& {Bastien},
  P. 2015, \aj, 149, 31

\bibitem[{{Keown} {et~al.}(2019){Keown}, {Di Francesco}, {Rosolowsky}, {Singh},
  {Figura}, {Kirk}, {Anderson}, {Chen}, {Elia}, {Friesen}, {Ginsburg},
  {Marston}, {Pezzuto}, {Schisano}, {Bontemps}, {Caselli}, {Liu}, {Longmore},
  {Motte}, {Myers}, {Offner}, {Sanhueza}, {Schneider}, {Stephens}, {Urquhart},
  \& {KEYSTONE Collaboration}}]{Keown+2019}
{Keown}, J., {Di Francesco}, J., {Rosolowsky}, E., {et~al.} 2019, \apj, 884, 4

\bibitem[{{Kharchenko} {et~al.}(2013){Kharchenko}, {Piskunov}, {Schilbach},
  {R{\"o}ser}, \& {Scholz}}]{Kharchenko+2013}
{Kharchenko}, N.~V., {Piskunov}, A.~E., {Schilbach}, E., {R{\"o}ser}, S., \&
  {Scholz}, R.~D. 2013, \aap, 558, A53

\bibitem[{{Kirchschlager} {et~al.}(2019){Kirchschlager}, {Bertrang}, \&
  {Flock}}]{Kirchschlager+2019}
{Kirchschlager}, F., {Bertrang}, G. H.~M., \& {Flock}, M. 2019, \mnras, 488,
  1211

\bibitem[{{Kirk} {et~al.}(2006){Kirk}, {Ward-Thompson}, \&
  {Crutcher}}]{Kirk+2006}
{Kirk}, J.~M., {Ward-Thompson}, D., \& {Crutcher}, R.~M. 2006, \mnras, 369,
  1445

\bibitem[{{K{\"o}nyves} {et~al.}(2015){K{\"o}nyves}, {Andr{\'e}},
  {Men'shchikov}, {Palmeirim}, {Arzoumanian}, {Schneider}, {Roy}, {Didelon},
  {Maury}, {Shimajiri}, {Di Francesco}, {Bontemps}, {Peretto}, {Benedettini},
  {Bernard}, {Elia}, {Griffin}, {Hill}, {Kirk}, {Ladjelate}, {Marsh}, {Martin},
  {Motte}, {Nguy{\^e}n Luong}, {Pezzuto}, {Roussel}, {Rygl}, {Sadavoy},
  {Schisano}, {Spinoglio}, {Ward-Thompson}, \& {White}}]{Konyves+2015}
{K{\"o}nyves}, V., {Andr{\'e}}, P., {Men'shchikov}, A., {et~al.} 2015, \aap,
  584, A91

\bibitem[{{K{\"o}nyves} {et~al.}(2020){K{\"o}nyves}, {Andr{\'e}},
  {Arzoumanian}, {Schneider}, {Men'shchikov}, {Bontemps}, {Ladjelate},
  {Didelon}, {Pezzuto}, {Benedettini}, {Bracco}, {Di Francesco}, {Goodwin},
  {Rygl}, {Shimajiri}, {Spinoglio}, {Ward-Thompson}, \& {White}}]{Konyves+2020}
{K{\"o}nyves}, V., {Andr{\'e}}, P., {Arzoumanian}, D., {et~al.} 2020, \aap,
  635, A34

\bibitem[{{Lai} {et~al.}(2002){Lai}, {Crutcher}, {Girart}, \& {Rao}}]{Lai+2002}
{Lai}, S.-P., {Crutcher}, R.~M., {Girart}, J.~M., \& {Rao}, R. 2002, \apj, 566,
  925

\bibitem[{{Lasker} {et~al.}(1990){Lasker}, {Sturch}, {McLean}, {Russell},
  {Jenkner}, \& {Shara}}]{Lasker+1990}
{Lasker}, B.~M., {Sturch}, C.~R., {McLean}, B.~J., {et~al.} 1990, \aj, 99, 2019

\bibitem[{{Lazarian} \& {Hoang}(2007)}]{Lazarian&Hoang2007}
{Lazarian}, A., \& {Hoang}, T. 2007, \mnras, 378, 910

\bibitem[{{Liu} {et~al.}(2018){Liu}, {Kim}, {Liu}, {Juvela}, {Zhang}, {Wu},
  {Li}, {Parsons}, {Soam}, {Goldsmith}, {Su}, {Tatematsu}, {Qin}, {Garay},
  {Hirota}, {Wouterloot}, {Chen}, {Evans}, {Graves}, {Kang}, {Li}, {Mardones},
  {Rawlings}, {Ren}, \& {Wang}}]{Liu+2018}
{Liu}, T., {Kim}, K.-T., {Liu}, S.-Y., {et~al.} 2018, \apjl, 869, L5

\bibitem[{{Lombardi} {et~al.}(2011){Lombardi}, {Alves}, \&
  {Lada}}]{Lombardi+2011}
{Lombardi}, M., {Alves}, J., \& {Lada}, C.~J. 2011, \aap, 535, A16

\bibitem[{{Mairs} {et~al.}(2015){Mairs}, {Johnstone}, {Kirk}, {Graves},
  {Buckle}, {Beaulieu}, {Berry}, {Broekhoven-Fiene}, {Currie}, {Fich},
  {Hatchell}, {Jenness}, {Mottram}, {Nutter}, {Pattle}, {Pineda}, {Salji}, {di
  Francesco}, {Hogerheijde}, {Ward-Thompson}, \& {JCMT Gould Belt survey
  Team}}]{Mairs+2015}
{Mairs}, S., {Johnstone}, D., {Kirk}, H., {et~al.} 2015, \mnras, 454, 2557

\bibitem[{{Martins} {et~al.}(2012){Martins}, {Mahy}, {Hillier}, \&
  {Rauw}}]{Martins+2012}
{Martins}, F., {Mahy}, L., {Hillier}, D.~J., \& {Rauw}, G. 2012, \aap, 538, A39

\bibitem[{{Matthews} {et~al.}(2009){Matthews}, {McPhee}, {Fissel}, \&
  {Curran}}]{Matthews+2009}
{Matthews}, B.~C., {McPhee}, C.~A., {Fissel}, L.~M., \& {Curran}, R.~L. 2009,
  \apjs, 182, 143

\bibitem[{{Motte} {et~al.}(2010){Motte}, {Zavagno}, {Bontemps}, {Schneider},
  {Hennemann}, {di Francesco}, {Andr{\'e}}, {Saraceno}, {Griffin}, {Marston},
  {Ward-Thompson}, {White}, {Minier}, {Men'shchikov}, {Hill}, {Abergel},
  {Anderson}, {Aussel}, {Balog}, {Baluteau}, {Bernard}, {Cox}, {Csengeri},
  {Deharveng}, {Didelon}, {di Giorgio}, {Hargrave}, {Huang}, {Kirk}, {Leeks},
  {Li}, {Martin}, {Molinari}, {Nguyen-Luong}, {Olofsson}, {Persi}, {Peretto},
  {Pezzuto}, {Roussel}, {Russeil}, {Sadavoy}, {Sauvage}, {Sibthorpe},
  {Spinoglio}, {Testi}, {Teyssier}, {Vavrek}, {Wilson}, \&
  {Woodcraft}}]{Motte+2010}
{Motte}, F., {Zavagno}, A., {Bontemps}, S., {et~al.} 2010, \aap, 518, L77

\bibitem[{{Mu{\v{z}}i{\'c}} {et~al.}(2019){Mu{\v{z}}i{\'c}}, {Scholz},
  {Pe{\~n}a Ram{\'\i}rez}, {Jayawardhana}, {Sch{\"o}del}, {Geers}, {Cieza}, \&
  {Bayo}}]{Muzic+2019}
{Mu{\v{z}}i{\'c}}, K., {Scholz}, A., {Pe{\~n}a Ram{\'\i}rez}, K., {et~al.}
  2019, \apj, 881, 79

\bibitem[{{Ogura} \& {Ishida}(1981)}]{Ogura+Ishida1981}
{Ogura}, K., \& {Ishida}, K. 1981, \pasj, 33, 149

\bibitem[{{Ostriker} {et~al.}(2001){Ostriker}, {Stone}, \&
  {Gammie}}]{Ostriker+2001}
{Ostriker}, E.~C., {Stone}, J.~M., \& {Gammie}, C.~F. 2001, \apj, 546, 980

\bibitem[{{Park} \& {Sung}(2002)}]{Park+Sung2002}
{Park}, B.-G., \& {Sung}, H. 2002, \aj, 123, 892

\bibitem[{{Pattle} \& {Fissel}(2019)}]{PattleFissel2019}
{Pattle}, K., \& {Fissel}, L. 2019, Frontiers in Astronomy and Space Sciences,
  6, 15

\bibitem[{{Pattle} {et~al.}(2015){Pattle}, {Ward-Thompson}, {Kirk}, {White},
  {Drabek-Maunder}, {Buckle}, {Beaulieu}, {Berry}, {Broekhoven-Fiene},
  {Currie}, {Fich}, {Hatchell}, {Kirk}, {Jenness}, {Johnstone}, {Mottram},
  {Nutter}, {Pineda}, {Quinn}, {Salji}, {Tisi}, {Walker-Smith}, {di Francesco},
  {Hogerheijde}, {Andr{\'e}}, {Bastien}, {Bresnahan}, {Butner}, {Chen},
  {Chrysostomou}, {Coude}, {Davis}, {Duarte-Cabral}, {Fiege}, {Friberg},
  {Friesen}, {Fuller}, {Graves}, {Greaves}, {Gregson}, {Griffin}, {Holland},
  {Joncas}, {Knee}, {K{\"o}nyves}, {Mairs}, {Marsh}, {Matthews},
  {Moriarty-Schieven}, {Rawlings}, {Richer}, {Robertson}, {Rosolowsky},
  {Rumble}, {Sadavoy}, {Spinoglio}, {Thomas}, {Tothill}, {Viti}, {Wouterloot},
  {Yates}, \& {Zhu}}]{Pattle+2015}
{Pattle}, K., {Ward-Thompson}, D., {Kirk}, J.~M., {et~al.} 2015, \mnras, 450,
  1094

\bibitem[{{Pattle} {et~al.}(2017){Pattle}, {Ward-Thompson}, {Berry},
  {Hatchell}, {Chen}, {Pon}, {Koch}, {Kwon}, {Kim}, {Bastien}, {Cho},
  {Coud{\'e}}, {Di Francesco}, {Fuller}, {Furuya}, {Graves}, {Johnstone},
  {Kirk}, {Kwon}, {Lee}, {Matthews}, {Mottram}, {Parsons}, {Sadavoy},
  {Shinnaga}, {Soam}, {Hasegawa}, {Lai}, {Qiu}, \& {Friberg}}]{Pattle+2017}
{Pattle}, K., {Ward-Thompson}, D., {Berry}, D., {et~al.} 2017, \apj, 846, 122

\bibitem[{{Pattle} {et~al.}(2019){Pattle}, {Lai}, {Hasegawa}, {Wang}, {Furuya},
  {Ward-Thompson}, {Bastien}, {Coud{\'e}}, {Eswaraiah}, {Fanciullo}, {di
  Francesco}, {Hoang}, {Kim}, {Kwon}, {Lee}, {Liu}, {Liu}, {Matsumura},
  {Onaka}, {Sadavoy}, \& {Soam}}]{Pattle+2019a}
{Pattle}, K., {Lai}, S.-P., {Hasegawa}, T., {et~al.} 2019, \apj, 880, 27

\bibitem[{{Pattle} {et~al.}(2020{\natexlab{a}}){Pattle}, {Lai}, {Di Francesco},
  {Sadavoy}, {Ward-Thompson}, {Johnstone}, {Hoang}, {Arzoumanian}, {Bastien},
  {Bourke}, {Coud{\'e}}, {Doi}, {Eswaraiah}, {Fanciullo}, {Furuya}, {Hwang},
  {Hull}, {Kang}, {Kim}, {Kirchschlager}, {Kwon}, {Kwon}, {Lee}, {Liu},
  {Redman}, {Soam}, {Tahani}, {Tamura}, \& {Tang}}]{Pattle+2020b}
{Pattle}, K., {Lai}, S.-P., {Di Francesco}, J., {et~al.} 2020{\natexlab{a}},
  arXiv e-prints, arXiv:2011.09765

\bibitem[{{Pattle} {et~al.}(2020{\natexlab{b}}){Pattle}, {Lai}, {Wright},
  {Coud{\'e}}, {Plambeck}, {Hoang}, {Tang}, {Bastien}, {Eswaraiah}, {Furuya},
  {Hwang}, {Inutsuka}, {Kim}, {Kirchschlager}, {Kwon}, {Lee}, {Liu}, {Lyo},
  {Ohashi}, {Rawlings}, {Tahani}, {Tamura}, {Soam}, {Wang}, \&
  {Ward-Thompson}}]{Pattle+2020a}
{Pattle}, K., {Lai}, S.-P., {Wright}, M., {et~al.} 2020{\natexlab{b}}, arXiv
  e-prints, arXiv:2009.14758

\bibitem[{{P{\'e}rez}(1991)}]{Perez1991}
{P{\'e}rez}, M.~R. 1991, \rmxaa, 22, 99

\bibitem[{{Perez} {et~al.}(1987){Perez}, {The}, \& {Westerlund}}]{Perez+1987}
{Perez}, M.~R., {The}, P.~S., \& {Westerlund}, B.~E. 1987, \pasp, 99, 1050

\bibitem[{{Phelps} \& {Lada}(1997)}]{Phelps+Lada1997}
{Phelps}, R.~L., \& {Lada}, E.~A. 1997, \apj, 477, 176

\bibitem[{{Pillai} {et~al.}(2015){Pillai}, {Kauffmann}, {Tan}, {Goldsmith},
  {Carey}, \& {Menten}}]{Pillai+2015}
{Pillai}, T., {Kauffmann}, J., {Tan}, J.~C., {et~al.} 2015, \apj, 799, 74

\bibitem[{{Planck Collaboration XXXIV} {et~al.}(2016){Planck Collaboration
  XXXIV}, {Aghanim}, {Alves}, {Arnaud}, {Arzoumanian}, {Aumont}, {Baccigalupi},
  {Banday}, {Barreiro}, {Bartolo}, {Battaner}, {Benabed}, {Benoit-L{\'e}vy},
  {Bernard}, {Bersanelli}, {Bielewicz}, {Bonaldi}, {Bonavera}, {Bond},
  {Borrill}, {Bouchet}, {Boulanger}, {Bracco}, {Burigana}, {Calabrese},
  {Cardoso}, {Catalano}, {Chamballu}, {Chiang}, {Christensen}, {Colombi},
  {Colombo}, {Combet}, {Couchot}, {Crill}, {Curto}, {Cuttaia}, {Danese},
  {Davies}, {Davis}, {de Bernardis}, {de Rosa}, {de Zotti}, {Delabrouille},
  {Dickinson}, {Diego}, {Dole}, {Donzelli}, {Dor{\'e}}, {Douspis}, {Ducout},
  {Dupac}, {Efstathiou}, {Elsner}, {En{\ss}lin}, {Eriksen}, {Falgarone},
  {Ferri{\`e}re}, {Finelli}, {Forni}, {Frailis}, {Fraisse}, {Franceschi},
  {Frejsel}, {Galeotta}, {Galli}, {Ganga}, {Ghosh}, {Giard}, {Gjerl{\o}w},
  {Gonz{\'a}lez-Nuevo}, {G{\'o}rski}, {Gregorio}, {Gruppuso}, {Guillet},
  {Hansen}, {Hanson}, {Harrison}, {Henrot-Versill{\'e}}, {Herranz},
  {Hildebrandt}, {Hivon}, {Hobson}, {Holmes}, {Hornstrup}, {Hovest},
  {Huffenberger}, {Hurier}, {Jaffe}, {Jaffe}, {Jewell}, {Juvela}, {Keskitalo},
  {Kisner}, {Knoche}, {Kunz}, {Kurki-Suonio}, {Lagache}, {Lamarre}, {Lasenby},
  {Lattanzi}, {Lawrence}, {Leonardi}, {Levrier}, {Liguori}, {Lilje},
  {Linden-V{\o}rnle}, {L{\'o}pez-Caniego}, {Lubin}, {Mac{\'\i}as-P{\'e}rez},
  {Maffei}, {Maino}, {Mand olesi}, {Mangilli}, {Maris}, {Martin},
  {Mart{\'\i}nez-Gonz{\'a}lez}, {Masi}, {Matarrese}, {Melchiorri}, {Mendes},
  {Mennella}, {Migliaccio}, {Miville-Desch{\^e}nes}, {Moneti}, {Montier},
  {Morgante}, {Mortlock}, {Moss}, {Munshi}, {Murphy}, {Naselsky}, {Nati},
  {Natoli}, {Netterfield}, {Noviello}, {Novikov}, {Novikov}, {Oppermann},
  {Pagano}, {Pajot}, {Paladini}, {Paoletti}, {Pasian}, {Patanchon},
  {Perdereau}, {Pettorino}, {Piacentini}, {Piat}, {Pietrobon}, {Plaszczynski},
  {Pointecouteau}, {Polenta}, {Ponthieu}, {Pratt}, {Pr{\'e}zeau}, {Prunet},
  {Puget}, {Rebolo}, {Reinecke}, {Remazeilles}, {Renault}, {Renzi},
  {Ristorcelli}, {Rocha}, {Rosset}, {Rossetti}, {Roudier},
  {Rubi{\~n}o-Mart{\'\i}n}, {Rusholme}, {Sandri}, {Santos}, {Savelainen},
  {Savini}, {Scott}, {Soler}, {Spencer}, {Stolyarov}, {Sutton}, {Suur-Uski},
  {Sygnet}, {Tauber}, {Terenzi}, {Toffolatti}, {Tomasi}, {Tristram}, {Tucci},
  {Tuovinen}, {Valenziano}, {Valiviita}, {Van Tent}, {Vielva}, {Villa}, {Wade},
  {Wandelt}, {Wehus}, {Wiesemeyer}, {Yvon}, {Zacchei}, \&
  {Zonca}}]{PlanckCollabXXXIV+2016}
{Planck Collaboration XXXIV}, {Aghanim}, N., {Alves}, M.~I.~R., {et~al.} 2016,
  \aap, 586, A137

\bibitem[{{Planck Collaboration XXXV} {et~al.}(2016){Planck Collaboration
  XXXV}, {Ade}, {Aghanim}, {Alves}, {Arnaud}, {Arzoumanian}, {Ashdown},
  {Aumont}, {Baccigalupi}, {Band ay}, {Barreiro}, {Bartolo}, {Battaner},
  {Benabed}, {Beno{\^\i}t}, {Benoit-L{\'e}vy}, {Bernard}, {Bersanelli},
  {Bielewicz}, {Bock}, {Bonavera}, {Bond}, {Borrill}, {Bouchet}, {Boulanger},
  {Bracco}, {Burigana}, {Calabrese}, {Cardoso}, {Catalano}, {Chiang},
  {Christensen}, {Colombo}, {Combet}, {Couchot}, {Crill}, {Curto}, {Cuttaia},
  {Danese}, {Davies}, {Davis}, {de Bernardis}, {de Rosa}, {de Zotti},
  {Delabrouille}, {Dickinson}, {Diego}, {Dole}, {Donzelli}, {Dor{\'e}},
  {Douspis}, {Ducout}, {Dupac}, {Efstathiou}, {Elsner}, {En{\ss}lin},
  {Eriksen}, {Falceta-Gon{\c{c}}alves}, {Falgarone}, {Ferri{\`e}re}, {Finelli},
  {Forni}, {Frailis}, {Fraisse}, {Franceschi}, {Frejsel}, {Galeotta}, {Galli},
  {Ganga}, {Ghosh}, {Giard}, {Gjerl{\o}w}, {Gonz{\'a}lez-Nuevo}, {G{\'o}rski},
  {Gregorio}, {Gruppuso}, {Gudmundsson}, {Guillet}, {Harrison}, {Helou},
  {Hennebelle}, {Henrot-Versill{\'e}}, {Hern{\'a}ndez-Monteagudo}, {Herranz},
  {Hildebrand t}, {Hivon}, {Holmes}, {Hornstrup}, {Huffenberger}, {Hurier},
  {Jaffe}, {Jaffe}, {Jones}, {Juvela}, {Keih{\"a}nen}, {Keskitalo}, {Kisner},
  {Knoche}, {Kunz}, {Kurki-Suonio}, {Lagache}, {Lamarre}, {Lasenby},
  {Lattanzi}, {Lawrence}, {Leonardi}, {Levrier}, {Liguori}, {Lilje},
  {Linden-V{\o}rnle}, {L{\'o}pez-Caniego}, {Lubin}, {Mac{\'\i}as-P{\'e}rez},
  {Maino}, {Mandolesi}, {Mangilli}, {Maris}, {Martin},
  {Mart{\'\i}nez-Gonz{\'a}lez}, {Masi}, {Matarrese}, {Melchiorri}, {Mendes},
  {Mennella}, {Migliaccio}, {Miville-Desch{\^e}nes}, {Moneti}, {Montier},
  {Morgante}, {Mortlock}, {Munshi}, {Murphy}, {Naselsky}, {Nati},
  {Netterfield}, {Noviello}, {Novikov}, {Novikov}, {Oppermann}, {Oxborrow},
  {Pagano}, {Pajot}, {Paladini}, {Paoletti}, {Pasian}, {Perotto}, {Pettorino},
  {Piacentini}, {Piat}, {Pierpaoli}, {Pietrobon}, {Plaszczynski},
  {Pointecouteau}, {Polenta}, {Ponthieu}, {Pratt}, {Prunet}, {Puget}, {Rachen},
  {Reinecke}, {Remazeilles}, {Renault}, {Renzi}, {Ristorcelli}, {Rocha},
  {Rossetti}, {Roudier}, {Rubi{\~n}o-Mart{\'\i}n}, {Rusholme}, {Sandri},
  {Santos}, {Savelainen}, {Savini}, {Scott}, {Soler}, {Stolyarov}, {Sudiwala},
  {Sutton}, {Suur-Uski}, {Sygnet}, {Tauber}, {Terenzi}, {Toffolatti}, {Tomasi},
  {Tristram}, {Tucci}, {Umana}, {Valenziano}, {Valiviita}, {Van Tent},
  {Vielva}, {Villa}, {Wade}, {Wandelt}, {Wehus}, {Ysard}, {Yvon}, \&
  {Zonca}}]{PlanckCollabXXXV+2016}
{Planck Collaboration XXXV}, {Ade}, P.~A.~R., {Aghanim}, N., {et~al.} 2016,
  \aap, 586, A138

\bibitem[{{Poulton} {et~al.}(2008){Poulton}, {Robitaille}, {Greaves},
  {Bonnell}, {Williams}, \& {Heyer}}]{Poulton+2008}
{Poulton}, C.~J., {Robitaille}, T.~P., {Greaves}, J.~S., {et~al.} 2008, \mnras,
  384, 1249

\bibitem[{{Povich} {et~al.}(2007){Povich}, {Stone}, {Churchwell}, {Zweibel},
  {Wolfire}, {Babler}, {Indebetouw}, {Meade}, \& {Whitney}}]{Povich+2007}
{Povich}, M.~S., {Stone}, J.~M., {Churchwell}, E., {et~al.} 2007, \apj, 660,
  346

\bibitem[{{Robitaille} \& {Bressert}(2012)}]{Robitaille+Bressert2012}
{Robitaille}, T., \& {Bressert}, E. 2012, {APLpy: Astronomical Plotting Library
  in Python}, , , ascl:1208.017

\bibitem[{{Rom{\'a}n-Z{\'u}{\~n}iga} {et~al.}(2008){Rom{\'a}n-Z{\'u}{\~n}iga},
  {Elston}, {Ferreira}, \& {Lada}}]{Roman-Zuniga+2008}
{Rom{\'a}n-Z{\'u}{\~n}iga}, C.~G., {Elston}, R., {Ferreira}, B., \& {Lada},
  E.~A. 2008, \apj, 672, 861

\bibitem[{{Rom{\'a}n-Z{\'u}{\~n}iga} \& {Lada}(2008)}]{Roman-Zuniga+Lada2008}
{Rom{\'a}n-Z{\'u}{\~n}iga}, C.~G., \& {Lada}, E.~A. 2008, {Star Formation in
  the Rosette Complex}, ed. B.~{Reipurth}, Vol.~4, 928

\bibitem[{{Roy} {et~al.}(2014){Roy}, {Andr{\'e}}, {Palmeirim}, {Attard},
  {K{\"o}nyves}, {Schneider}, {Peretto}, {Men'shchikov}, {Ward-Thompson},
  {Kirk}, {Griffin}, {Marsh}, {Abergel}, {Arzoumanian}, {Benedettini}, {Hill},
  {Motte}, {Nguyen Luong}, {Pezzuto}, {Rivera-Ingraham}, {Roussel}, {Rygl},
  {Spinoglio}, {Stamatellos}, \& {White}}]{Roy+2014}
{Roy}, A., {Andr{\'e}}, P., {Palmeirim}, P., {et~al.} 2014, \aap, 562, A138

\bibitem[{{Savage} {et~al.}(2013){Savage}, {Spangler}, \&
  {Fischer}}]{Savage+2013}
{Savage}, A.~H., {Spangler}, S.~R., \& {Fischer}, P.~D. 2013, \apj, 765, 42

\bibitem[{{Schneider} {et~al.}(2010){Schneider}, {Motte}, {Bontemps},
  {Hennemann}, {di Francesco}, {Andr{\'e}}, {Zavagno}, {Csengeri},
  {Men'shchikov}, {Abergel}, {Baluteau}, {Bernard}, {Cox}, {Didelon}, {di
  Giorgio}, {Gastaud}, {Griffin}, {Hargrave}, {Hill}, {Huang}, {Kirk},
  {K{\"o}nyves}, {Leeks}, {Li}, {Marston}, {Martin}, {Minier}, {Molinari},
  {Olofsson}, {Panuzzo}, {Persi}, {Pezzuto}, {Roussel}, {Russeil}, {Sadavoy},
  {Saraceno}, {Sauvage}, {Sibthorpe}, {Spinoglio}, {Testi}, {Teyssier},
  {Vavrek}, {Ward-Thompson}, {White}, {Wilson}, \&
  {Woodcraft}}]{Schneider+2010}
{Schneider}, N., {Motte}, F., {Bontemps}, S., {et~al.} 2010, \aap, 518, L83

\bibitem[{{Schneider} {et~al.}(2012){Schneider}, {Csengeri}, {Hennemann},
  {Motte}, {Didelon}, {Federrath}, {Bontemps}, {Di Francesco}, {Arzoumanian},
  {Minier}, {Andr{\'e}}, {Hill}, {Zavagno}, {Nguyen-Luong}, {Attard},
  {Bernard}, {Elia}, {Fallscheer}, {Griffin}, {Kirk}, {Klessen}, {K{\"o}nyves},
  {Martin}, {Men'shchikov}, {Palmeirim}, {Peretto}, {Pestalozzi}, {Russeil},
  {Sadavoy}, {Sousbie}, {Testi}, {Tremblin}, {Ward-Thompson}, \&
  {White}}]{Schneider+2012}
{Schneider}, N., {Csengeri}, T., {Hennemann}, M., {et~al.} 2012, \aap, 540, L11

\bibitem[{{Simpson} {et~al.}(2012){Simpson}, {Povich}, {Kendrew}, {Lintott},
  {Bressert}, {Arvidsson}, {Cyganowski}, {Maddison}, {Schawinski}, {Sherman},
  {Smith}, \& {Wolf-Chase}}]{Simpson+2012}
{Simpson}, R.~J., {Povich}, M.~S., {Kendrew}, S., {et~al.} 2012, \mnras, 424,
  2442

\bibitem[{{Soam} {et~al.}(2019){Soam}, {Liu}, {Andersson}, {Lee}, {Liu},
  {Juvela}, {Li}, {Goldsmith}, {Zhang}, {Koch}, {Kim}, {Qiu}, {Evans},
  {Johnstone}, {Thompson}, {Ward-Thompson}, {Di Francesco}, {Tang},
  {Montillaud}, {Kim}, {Mairs}, {Sanhueza}, {Kim}, {Berry}, {Gordon},
  {Tatematsu}, {Liu}, {Pattle}, {Eden}, {McGehee}, {Wang}, {Ristorcelli},
  {Graves}, {Alina}, {Lacaille}, {Montier}, {Park}, {Kwon}, {Chung},
  {Pelkonen}, {Micelotta}, {Saajasto}, \& {Fuller}}]{Soam+2019}
{Soam}, A., {Liu}, T., {Andersson}, B.~G., {et~al.} 2019, \apj, 883, 95

\bibitem[{{Townsley} {et~al.}(2003){Townsley}, {Feigelson}, {Montmerle},
  {Broos}, {Chu}, \& {Garmire}}]{Townsley+2003}
{Townsley}, L.~K., {Feigelson}, E.~D., {Montmerle}, T., {et~al.} 2003, \apj,
  593, 874

\bibitem[{{Tremblin} {et~al.}(2013){Tremblin}, {Minier}, {Schneider}, {Audit},
  {Hill}, {Didelon}, {Peretto}, {Arzoumanian}, {Motte}, {Zavagno}, {Bontemps},
  {Anderson}, {Andr{\'e}}, {Bernard}, {Csengeri}, {Di Francesco}, {Elia},
  {Hennemann}, {K{\"o}nyves}, {Marston}, {Nguyen Luong}, {Rivera-Ingraham},
  {Roussel}, {Sousbie}, {Spinoglio}, {White}, \& {Williams}}]{Tremblin+2013}
{Tremblin}, P., {Minier}, V., {Schneider}, N., {et~al.} 2013, \aap, 560, A19

\bibitem[{{Tremblin} {et~al.}(2014){Tremblin}, {Schneider}, {Minier},
  {Didelon}, {Hill}, {Anderson}, {Motte}, {Zavagno}, {Andr{\'e}},
  {Arzoumanian}, {Audit}, {Benedettini}, {Bontemps}, {Csengeri}, {Di
  Francesco}, {Giannini}, {Hennemann}, {Nguyen Luong}, {Marston}, {Peretto},
  {Rivera-Ingraham}, {Russeil}, {Rygl}, {Spinoglio}, \&
  {White}}]{Tremblin+2014}
{Tremblin}, P., {Schneider}, N., {Minier}, V., {et~al.} 2014, \aap, 564, A106

\bibitem[{{Wang} {et~al.}(2009){Wang}, {Feigelson}, {Townsley},
  {Rom{\'a}n-Z{\'u}{\~n}iga}, {Lada}, \& {Garmire}}]{Wang+2009}
{Wang}, J., {Feigelson}, E.~D., {Townsley}, L.~K., {et~al.} 2009, \apj, 696, 47

\bibitem[{{Ward-Thompson} {et~al.}(2017){Ward-Thompson}, {Pattle}, {Bastien},
  {Furuya}, {Kwon}, {Lai}, {Qiu}, {Berry}, {Choi}, {Coud{\'e}}, {Di Francesco},
  {Hoang}, {Franzmann}, {Friberg}, {Graves}, {Greaves}, {Houde}, {Johnstone},
  {Kirk}, {Koch}, {Kwon}, {Lee}, {Li}, {Matthews}, {Mottram}, {Parsons}, {Pon},
  {Rao}, {Rawlings}, {Shinnaga}, {Sadavoy}, {van Loo}, {Aso}, {Byun},
  {Eswaraiah}, {Chen}, {Chen}, {Chen}, {Ching}, {Cho}, {Chrysostomou}, {Chung},
  {Doi}, {Drabek-Maunder}, {Eyres}, {Fiege}, {Friesen}, {Fuller}, {Gledhill},
  {Griffin}, {Gu}, {Hasegawa}, {Hatchell}, {Hayashi}, {Holland}, {Inoue},
  {Inutsuka}, {Iwasaki}, {Jeong}, {Kang}, {Kang}, {Kang}, {Kawabata}, {Kemper},
  {Kim}, {Kim}, {Kim}, {Kim}, {Kim}, {Kim}, {Lacaille}, {Lee}, {Lee}, {Li},
  {Li}, {Liu}, {Liu}, {Liu}, {Liu}, {Lyo}, {Mairs}, {Matsumura},
  {Moriarty-Schieven}, {Nakamura}, {Nakanishi}, {Ohashi}, {Onaka}, {Peretto},
  {Pyo}, {Qian}, {Retter}, {Richer}, {Rigby}, {Robitaille}, {Savini}, {Scaife},
  {Soam}, {Tamura}, {Tang}, {Tomisaka}, {Wang}, {Wang}, {Whitworth}, {Yen},
  {Yoo}, {Yuan}, {Zhang}, {Zhang}, {Zhou}, {Zhu}, {Andr{\'e}}, {Dowell},
  {Falle}, \& {Tsukamoto}}]{Ward-Thompson+2017}
{Ward-Thompson}, D., {Pattle}, K., {Bastien}, P., {et~al.} 2017, \apj, 842, 66

\bibitem[{{Whittet} {et~al.}(2008){Whittet}, {Hough}, {Lazarian}, \&
  {Hoang}}]{Whittet+2008}
{Whittet}, D.~C.~B., {Hough}, J.~H., {Lazarian}, A., \& {Hoang}, T. 2008, \apj,
  674, 304

\bibitem[{{Williams} {et~al.}(1995){Williams}, {Blitz}, \&
  {Stark}}]{Williams+1995}
{Williams}, J.~P., {Blitz}, L., \& {Stark}, A.~A. 1995, \apj, 451, 252

\bibitem[{{Ybarra} {et~al.}(2013){Ybarra}, {Lada}, {Rom{\'a}n-Z{\'u}{\~n}iga},
  {Balog}, {Wang}, \& {Feigelson}}]{Ybarra+2013}
{Ybarra}, J.~E., {Lada}, E.~A., {Rom{\'a}n-Z{\'u}{\~n}iga}, C.~G., {et~al.}
  2013, \apj, 769, 140

\bibitem[{{Ybarra} \& {Phelps}(2004)}]{Ybarra+Phelps2004}
{Ybarra}, J.~E., \& {Phelps}, R.~L. 2004, \aj, 127, 3444

\end{thebibliography}
\bibliographystyle{aasjournal}



\end{document}